%% file: param.tex
\documentclass[11pt]{article}
\pdfoutput=1
\usepackage{amsmath,amsfonts,sint,epsfig,cite,graphics}
\usepackage{amssymb}
\usepackage{mathrsfs}  
\usepackage{macros_static}
  \renewcommand{\cahqet}{\cah{1}}
\usepackage{color}
\begin{document}

\input title.tex

\input s1.tex

\input s2.tex

\input s3.tex

\input s4.tex

\input s5.tex

\noindent
{\bf Acknowledgements.}
We would like to thank G. von Hippel, H. Wittig and 
in particular H. Simma for very useful comments on the manuscript. 
This work is supported by the Deutsche Forschungsgemeinschaft
in the SFB/TR~09
and by the European community
through EU Contract No.~MRTN-CT-2006-035482, ``FLAVIAnet''.
N.G. acknowledges financial support from the MICINN grant FPA2006-05807, the 
Comunidad Autónoma de Madrid programme HEPHACOS P-ESP-00346, and participates 
in the Consolider-Ingenio 2010 CPAN (CSD2007-00042). 
\begin{appendix}

\input a1.tex
\input a2.tex

\input a3.tex
\input a4.tex

\end{appendix}

\bibliographystyle{JHEP}   
\bibliography{refs}

\end{document}

%% file: title.tex
\begin{titlepage}

\begin{flushright}
\small{
IFT-UAM/CSIC-09-47\\
Edinburgh 2009/24\\
MKPH-T-09-18 \\
DESY 09-153 \\
SFB/CPP-10-01\\
LPT-Orsay/10-01\\}
\end{flushright}

\begin{center}
{\Large\bf
HQET at order $1/m$: I.
Non-perturbative parameters in the quenched approximation
}
\end{center}
\vskip 0.35cm
\vbox{
\centerline{
\epsfxsize=2.8 true cm
\epsfbox{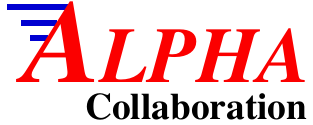}}
}
\vskip 0.1cm
\begin{center}
{
Beno\^it Blossier$^{\scriptscriptstyle a}$,
Michele Della Morte$^{\scriptscriptstyle b}$,
Nicolas Garron$^{\scriptscriptstyle c,d}$,
Rainer Sommer$^{\scriptscriptstyle e}$ 
}
\vskip 0.5cm
{
\vskip 2.0ex
$^{\scriptstyle a}$
Laboratoire de Physique Th\'eorique, B\^atiment 210, Universit\'e Paris XI,
F-91405 Orsay Cedex, France
\vskip 2.0ex
$^{\scriptstyle b}$
Institut f{\"u}r Kernphysik, University of Mainz,
D-55099 Mainz, Germany
\vskip 2.0ex
$^{\scriptstyle c}$
Dpto F\'isica Te\'orica and Instituto de F\'isica Te\'orica UAM/CSIC\\
Universidad Aut\'onoma de Madrid, Cantoblanco E-28049 Madrid, Spain
\vskip 2.0ex
$^{\scriptstyle d}$
SUPA, School of Physics and Astronomy, Univ. of Edinburgh, EH9 3JZ, UK.
\vskip 2.0ex
$^{\scriptstyle e}$
NIC, DESY,
Platanenallee 6, 15738 Zeuthen,  Germany
\vskip 2.0ex
}
\vskip 0.5cm
{\bf Abstract}
\vskip 0.1ex
\end{center}
We determine non-perturbatively the parameters 
of the lattice HQET Lagrangian and those of
the time component of the heavy-light axial-vector current
in the quenched approximation.
The HQET expansion includes terms of order $\minv$.
Our results allow to compute, for example, 
the heavy-light spectrum and B-meson decay constants
in the static approximation and to order $\minv$ in HQET.
The determination of the parameters is separated into universal and
non-universal parts. The universal results can be used to
determine the parameters for various discretizations. The
computation reported in this paper uses 
the plaquette gauge action and the ``HYP1/2''
action for the b-quark described by HQET. The parameters
of the current also depend on the light-quark action,
for which we choose non-perturbatively $O(a)$-improved Wilson fermions.


\vskip 2.0ex

\eject
\vfill
\eject

\end{titlepage}

%% file: s1.tex
\section{Introduction}

Heavy quark effective theory (HQET) was developed already quite a while ago
\cite{stat:eichhill1,stat:symm1,stat:symm3,Eichten:1990vp, hqet:cont3}.
Still it is of considerable interest today,  
primarily for two reasons. First it describes the asymptotic expansion
of QCD observables in the limit of a large quark mass,
in particular the mass of the b-quark, $\mbeauty$. For this
to be true, 
the observables have to be in the proper kinematical region.
But in such
a region, we expect the expansion in $\minv$  to be valid also 
non-perturbatively {\em if}
the parameters in the effective theory are determined 
non-perturbatively. Understanding QCD then includes understanding
the HQET limit. Steps towards establishing the expected equivalence 
on the non-perturbative level\footnote{Note that to our knowledge a 
proof of the renormalizability of HQET to all orders 
of perturbation theory has not yet been given.}
 have
already been carried out~\cite{hqet:pap2,lat08:patrick}, but we intend
to go much further, in particular through a complete treatment of $\minv$
corrections.

Second, flavour physics has become a precision field. 
No flavour physics observable has shown evidence for
physics beyond the Standard Model with the presently available
precision. In many cases the limiting part is the theoretical uncertainty, not
the experimental one~\cite{CKM08}. Reliable lattice computations are 
needed to make progress, but 
heavy quarks are difficult due to $\rmO((a \mbeauty)^n)$
discretization errors, where $a$ is the lattice spacing. 
In particular, if $a \mbeauty$ is too large the 
expansion in $a$ will break down altogether~\cite{zastat:pap2,fds:quenched2}
\footnote{The second one of these references shows the difficulty 
even for a charm quark. For more general discussions and
more references we refer the reader to 
reviews~\cite{lat03:kronfeld,qcd06:garron,lat07:michele,lat08:gamiz,lat09:aubin}.}.
Instead, in HQET the cutoff effects are 
$\rmO((a \Lambda_{\rm QCD})^n)$; they are thus much more manageable. 

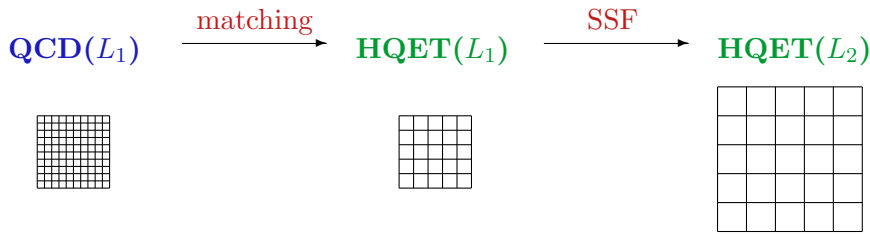
\begin{figure}[!htb]
\begin{center}
\hspace{-0.5cm}

\definecolor{r1}{rgb}{.75, .10, .10}
\definecolor{b1}{rgb}{.10, .10, .75}
\definecolor{g1}{rgb}{0.0, .60, .20}

  \setlength{\unitlength}{0.5em}
  \begin{picture}(60,5)
     \put(0,0) { \color{b1} \bf QCD($L_1$) }
     \put(12,1){ \vector(1,0){10} }
     \put(13,2){ \color{r1} matching }
     \put(24,0){ \color{g1} \bf HQET($L_1$) }
     \put(37,1){ \vector(1,0){10} }
     \put(40,2){ \color{r1} SSF }
     \put(49,0){ \color{g1} \bf HQET($L_2$) }
     \multiput(2,-4)  (0.5,0)  {11}{ \line(0,-1){5} }
     \multiput(2,-4)  (0,-0.5) {11}{ \line(1,0){5} }
     \multiput(27,-4) (1,0)  {6}{ \line(0,-1){5} }
     \multiput(27,-4) (0,-1) {6}{ \line(1,0){5} }
     \multiput(49,-2) (2,0)  {6}{ \line(0,-1){10} }
     \multiput(49,-2) (0,-2) {6}{ \line(1,0){10} }
  \end{picture}
  \vspace*{6em}
\end{center}
\vspace{-0.7cm}
\caption[]{\footnotesize 
The strategy.
}
\label{f:sketch}
\end{figure}

Here we present the determination of the parameters 
of the effective theory in a formulation where all
power divergences are subtracted non-perturbatively 
through a matching of QCD and HQET in a small volume \cite{hqet:pap1}
with \SF boundary conditions. As illustrated in \fig{f:sketch},
we then continue to larger lattice spacings through a
step scaling method. \Sect{s:param} explains the details,
in particular how lattice spacing errors are removed in each step.
The work of \cite{hqet:pap4} is here extended by a determination
of all parameters in the action as well as those for the 
weak current $A_0$ including the terms of order $\minv$. 

Our numerical implementation is done in the quenched approximation.
It represents a very useful test-laboratory, in particular
for the first motivation presented above: the qualitative features
of the $\minv$ expansion will not depend on the fact that the
light quarks are quenched. Applications of the parameters 
computed in this paper require the determination of
energies and matrix elements in a large volume and
will appear in separate papers, but already here we learn
interesting lessons about the asymptotic convergence of the 
expansion. In the following section we recall the basic formulation of
HQET, with a primary focus on defining the parameters of the 
theory and how they enter the computation of the spectrum and 
matrix elements. Section 3 gives a short but precise account of the 
strategy for the computation of the parameters. The technical Sections 4.1-4.2 
discuss the numerical details of the different intermediate steps while 
Section 4.3 presents our main results for the HQET parameters.

%% file: s2.tex
\section{HQET at order  $\minv$}
In this section we define HQET including terms
of order  $\minv$, in particular the parameters
of the theory. We then show the expansion of a
few observables as examples how the parameters 
can be used. 
We choose to regularise the theory on the lattice 
although our approach is in principle independent of 
the specific regularisation.
Almost all formulae are established 
in~\cite{hqet:pap4}. They are repeated here for the benefit 
of the reader to keep the paper self-contained,
but for details, such as the exact consequence of 
spin symmetry, the reader is referred to~\cite{hqet:pap4}.
Terms of order $\minv^2$ are dropped without notice.

\subsection{Lagrangian}
The HQET Lagrangian,
\bes
\lag{HQET}(x) &=&  \lag{stat}(x) - \omegakin\Okin(x)
        - \omegaspin\Ospin(x)  \,,
\ees
consists of the lowest order (static) term,
\be
\label{e:statact}
\lag{stat}(x) =
\heavyb(x) \,D_0\, \heavy(x) \;,
\ee
and the first order corrections
\bes
  \Okin(x) &=& \heavyb(x){\bf D}^2\heavy(x) \,,\quad
  \Ospin(x) = \heavyb(x){\boldsymbol\sigma}\!\cdot\!{\bf B}\heavy(x)\,.
\ees
We use the
backward covariant derivative $D_0$ as in~\cite{stat:actpaper}
and
the 4-component  heavy quark field subject to the constraints
$P_+ \heavy = \heavy \,, \; \heavyb P_+=\heavyb$ with
$P_+={(1+\gamma_0)/{2}}$ and the discretized version
$
 {\boldsymbol\sigma}\!\cdot\!{\bf B} = \sum_{k,j}\sigma_{kj}\widehat{F}_{kj}/(2i)\,,
$
where $\sigma_{kj}$ and the lattice (clover) field tensor $\widehat{F}$ are defined
in~\cite{zastat:pap1}. The kinetic term ${\bf D}^2$ is represented
by the nearest neighbour covariant 3-d Laplacian. The normalization
is such that the {\em classical} values of the coefficients
are $\omegakin=\omegaspin=1/(2\mbeauty)$. A bare mass $\mhbare$ has to
be added to the energy levels (e.g. the B-meson
mass) computed with this Lagrangian to obtain the QCD ones (up to $\minv^2$).
At the classical level it is $\mbeauty$, but in the quantized 
theory, it has to further compensate a power divergence. 

\subsection{Weak axial current}
For many applications, the weak, left handed, heavy-light current is needed.
We here consider just the time component of the axial-vector part, $A_0$. 
The other components can be treated analogously,
but so far we restricted ourselves to $A_0$, which is sufficient
for a computation of pseudoscalar decay constants.

At the lowest order the current is form-identical to the 
relativistic one. At first order it is corrected by two
composite fields of dimension four. The explicit form is
\bes
 \label{e:ahqet}
 \Ahqet(x)&=& \zahqet\,[\Astat(x)+  \sum_{i=1}^2\cah{i}\Ah{i}(x)]\,, \\
 \Ah{1}(x) &=& \lightb(x){1\over2}
            \gamma_5\gamma_i(\nabsym{i}-\lnabsym{i})\heavy(x)\,,
 \label{e:dahqet}
\\
 \Ah{2}(x) &=& -\drvsym{i}\,\Ahi(x)\,,\quad \Ahi(x)=\lightb(x)\gamma_i\gamma_5\heavy(x)\,,
\ees
where all derivatives are taken to be the symmetric 
nearest neighbor ones,
\bes
  \drvsym{i} = \frac12(\drv{i}+\drvstar{i})\,,\quad
  \lnabsym{i} = \frac12(\lnab{i}+\lnabstar{i})\,,\quad
  \nabsym{i} = \frac12(\nab{i}+\nabstar{i})\,.
\ees
By considering $\zahqet$ to be a function of $m_\mrm{l}/\mbeauty$ 
with the light quark masses $m_\mrm{l}$, the above set of operators
is complete after using the symmetries and the equations of motion.
Since $m_\mrm{l} \lll \mbeauty$ and such effects are further reduced
by a factor of the coupling constant $\alpha(\mbeauty)$ we ignore
this dependence and determine $\zahqet$ with the light quark mass
set to zero.
Note that $\Ah{2}(x)$ does not contribute to 
correlation functions and matrix elements at 
vanishing space-momentum. It is not needed for the
computation of decay constants and has not been written down in
\cite{hqet:pap4}. We will also not determine its coefficient $\cah{2}$ here. 

A short note on discretization errors is useful at this point. 
Remaining in the static approximation ($\omegakin=\omegaspin=0$),
the Lagrangian has been shown to be automatically $\rmO(a)$ 
improved by studying its Symanzik expansion.
Therefore, all energy levels are as well~\cite{zastat:pap1}. 
The zero space-momentum matrix elements of the current $\Ahqet$ 
are $\rmO(a)$ improved in static approximation if one
sets $\cah{1}=a\castat$~\cite{zastat:pap1,stat:actpaper}.~\footnote{
At zero space momentum, partial summation can be used
to bring $\Ah{2}$ into the form used in \cite{zastat:pap1}.}
Including the $\minv$ terms, linear terms in $a$ remain absent,
except for those accompanied by a factor $\minv$~\cite{hqet:pap1}. 
So leading discretization errors are $\rmO(a/\mbeauty)$ and $\rmO(a^2)$.

\subsection{Observables}
\subsubsection{Correlation functions}
Observables of interest are obtained from 
Euclidean correlation functions in large volume. 
As an illustration we consider the QCD correlator
\bes \label{e_caa}
  \caa(x_0) = \za^2 a^3\sum_{\vecx} \Big\langle A_0(x)  (A_0(0))^{\dagger}
              \Big\rangle
\ees
of the heavy-light axial current in QCD,
$A_\mu=\lightb\gamma_\mu\gamma_5\psi_\beauty$ (and
$A_\mu^\dagger=-\psibar_\beauty\gamma_\mu\gamma_5\light$). It is normalized 
by $\za$ to 
satisfy the chiral Ward identities~\cite{boch,impr:pap4}. 
Ignoring renormalization for a moment, 
an expansion in $\minv$ reads 
\bes
   \caa(x_0) &=& \caa^\stat(x_0)+\caa^\first(x_0) = 
   \caa^\stat(x_0) \,[1 + \Raa^\first(x_0)]
\ees
where the notation indicates
$   
    \caa^\first=\rmO(\minv)
$.
In HQET the path integral {\em weight is expanded in $\minv$}
and one obtains the fully renormalized expansion
\bes
   \label{e:caahqet0}
   \caa(x_0) &=&  \rme^{-\mhbare x_0} (\zahqet)^2 \,\Big[
        \caa^\mrm{stat}(x_0)+\omegakin\caa^\mrm{kin}(x_0)+
        \omegaspin\caa^\mrm{spin}(x_0) \nonumber \\
        && \qquad \qquad \qquad\qquad\;+ 
         \cahqet[\cdaa^\mrm{stat}(x_0)+\cada^\mrm{stat}(x_0)]
        \Big]\,
        \label{e:caahqet} \\
        &\equiv&  \rme^{-\mhbare x_0} (\zahqet)^2 \,
        \caa^\mrm{stat}(x_0) \times \\ \nonumber
        && \qquad \qquad\;\Big[1+ \omegakin\Raa^\mrm{kin}(x_0)+
        \omegaspin\Raa^\mrm{spin}(x_0) +
         \cahqet\,R_{\delta A}(x_0)
        \Big]\,       
\ees
in terms of bare static expectation values, for example
\bes \label{e:caakinr}
 \caa^\mrm{kin}(x_0) = a^7 \sum_{\vecx,\, z}\Big\langle
    \Astat(x)\, (\Astat(0))^\dagger  \,\Okin(z)\Big\rangle_\mrm{stat} \,.
\ees
\subsubsection{Spectrum and matrix elements \label{s:spect}}
\newcommand{\ampl}{{\cal A}}
The spectral representation yields the large time behaviour in QCD
\bes
   \caa(x_0) = 
    \ampl^2 
    \,\rme^{-m_\mrm{B}x_0 }\,
   [\,1 + \rmO(\rme^{-\Delta\,x_0 })\,]\,,\label{e:caaspect} 
\ees
where $\Delta$ is a gap in this channel ( $\approx 2 \mpi$ in large {volume})
and  
$\ampl= \langle B(\vecp=0) | A_0(0) |0\rangle$ in the non-relativistic 
normalization of states  
$\langle B(\vecp) |  B(\vecp')\rangle =2\delta(\vecp-\vecp')$ . 
Of course  $\ampl=f_\mrm{B} \sqrt{m_\mrm{B}/2}$ is a phenomenologically 
interesting quantity. Again a naive expansion reads
\bes
   \caa(x_0) &=& (\ampl^\stat)^2\, \rme^{-m_\mrm{B}^\stat x_0 } 
    [1 + 2 \ampl^\first 
    - m_\mrm{B}^\first x_0]\,.
   \label{e:caaexp}  
\ees
Using the transfer matrix of the static theory,
the HQET correlators (static and beyond) are
easily analysed. The result is 
\bes
   \caa^\mrm{stat}(x_0) = (\ampl^\stat)^2\,\rme^{-\Estat x_0 }\,
   [\,1 + \rmO(\rme^{-\Delta^\stat\,x_0 })\,]
\ees
for the static term and for example
\bes
   \Raa^\mrm{kin}(x_0) = 2\ampl^\mrm{kin} -x_0\Ekin + 
   \rmO(x_0\,\rme^{-\Delta^\stat\,x_0 })\,
\ees
for the kinetic correction.\footnote{
An explicit expression for $\Ekin$ is
\bes
 \label{e:ekin}
    \Ekin &=& - {1\over 2L^3}\langle B | a^3\sum_{\vecz} \Okin(0,\vecz)| B \rangle_\mrm{stat} = - {1\over 2}\langle B | \Okin(0)| B \rangle_\mrm{stat}   
  \,,
\ees
where the state $| B \rangle$ is an eigenstate of the 
static transfer matrix.
}

A comparison to \eq{e:caaspect} yields
\bes
   \mB &=& \mhbare + \Estat + \omegakin \Ekin +  \omegaspin \Espin \,,
   \\ \label{e:expandfb}
   \ln(\ampl\,r_0^{3/2})  &=&  \ln(\zahqet) + \ln(\ampl^\stat\,r_0^{3/2}) 
   + \cahqet  \ampl^{\delta A} + \omegakin \ampl^\mrm{kin} 
            + \omegaspin \ampl^\mrm{spin}\,, \nonumber\\[-1ex]
\ees
where $r_0$ is an arbitrary length-scale of the theory.
Completely analogous formulae hold for the excited state masses 
and matrix elements. The necessary bare quantities
such as $\Ekin$ can be efficiently computed with the 
generalized eigenvalue method~\cite{gevp:pap}.
We have here chosen to write the expansion of 
$\ln(\ampl\,r_0^{3/2})$ instead of $\ampl$ itself, since in this way one 
explicitly avoids terms quadratic in $\minv$, while e.g.
in \eq{e:caahqet} terms such as 
$Z_\mrm{A}^{(\minv)} \omegakin \caa^\mrm{kin}$ {\em are understood 
to be dropped} to remain consistently in order $\minv$. This
rule is necessary for a correct renormalization of the theory.

\subsection{Properties of non-perturbative HQET parameters \label{s:properties}}
Before entering the presentation of a computation of
the HQET parameters, we here mention some of their properties.

\subsubsection{Scheme dependence \label{s:ambig}}
Since we are working in the framework of an {\em effective field theory},
which has non-trivial renormalization, the parameters 
have to be determined by a matching performed at finite value of the 
expansion parameter $\minv$. The truncation of the effective
theory then introduces an ambiguity of order of the 
left out terms. 

As a result the parameters such as $\omegakin$ and $\zahqet$
have a dependence on
the choice of the matching condition. 
This is a scheme dependence, analogous to the one of 
the usual perturbative expansion in the renormalized QCD coupling. For an explicit
example consider $\ln(\zahqet)$. Evaluated in the static approximation
we will denote it by $\ln(Z_\mrm{A}^\stat)$ and including all 
$\minv$ terms it is denoted as 
$\ln(\zahqet)=\ln(Z_\mrm{A}^\stat)+\ln(Z_\mrm{A}^{(\minv)})$.
The scheme dependence of $\ln(\zahqet)$ is then of order $\minv^2$,
while the one of $\ln(Z_\mrm{A}^\stat)$ and $\ln(Z_\mrm{A}^{(\minv)})$ individually is
of order $\minv$. Note that in all this discussion we are working
non-perturbatively in the QCD coupling but order by order in $\minv$.

\subsubsection{Difference to previous renormalizations
 in the static approximation \label{s:diff}}
In order to avoid a misunderstanding we also point out the 
difference to the renormalization carried out for example
in \cite{zastat:pap3,hqet:pap3}.
As long as one is working just in the static approximation,
the weak currents do not mix with operators of different dimensions. 
It is then also consistent to renormalize them perturbatively.
Both to avoid ambiguities in renormalization schemes and to be
able to profit from the high order continuum perturbative 
results~\cite{ChetGrozin,hqet:match_3loop},
it is advantageous to first introduce the RGI current, 
\bes
 \Argi(x)= \zastatrgi\,\Astat(x)\,.
\ees
The renormalization constant $\zastatrgi$ has been determined non-perturbatively
in \cite{zastat:pap3} for the quenched case, and in \cite{zastat:nf2}
for two flavours of dynamical quarks.

However, here we determine $\zastat$ 
by a non-perturbative matching to QCD, and it is
different. In fact the correspondence is 
\bes
  \zastat = \zastatrgi \Cps(\Mbeauty/\Lambda)\,,
\ees
in terms of the matching function $\Cps$ introduced in
\cite{hqet:pap3}.  In the ``old'' strategy of for example
\cite{hqet:pap3,DellaMorte:2007ij}, functions such as $\Cps$
are determined from high-order perturbation
theory, while here we evaluate the full factor $\zastat$ 
non-perturbatively.

%% file: s3.tex
\section{Computation of HQET parameters}
\label{s:param}
\input{tables/hqet_param_def.tex}

\begin{figure}[!htb]
\begin{center}
\hspace{-0.5cm}
\centerline{\includegraphics[width=\textwidth]{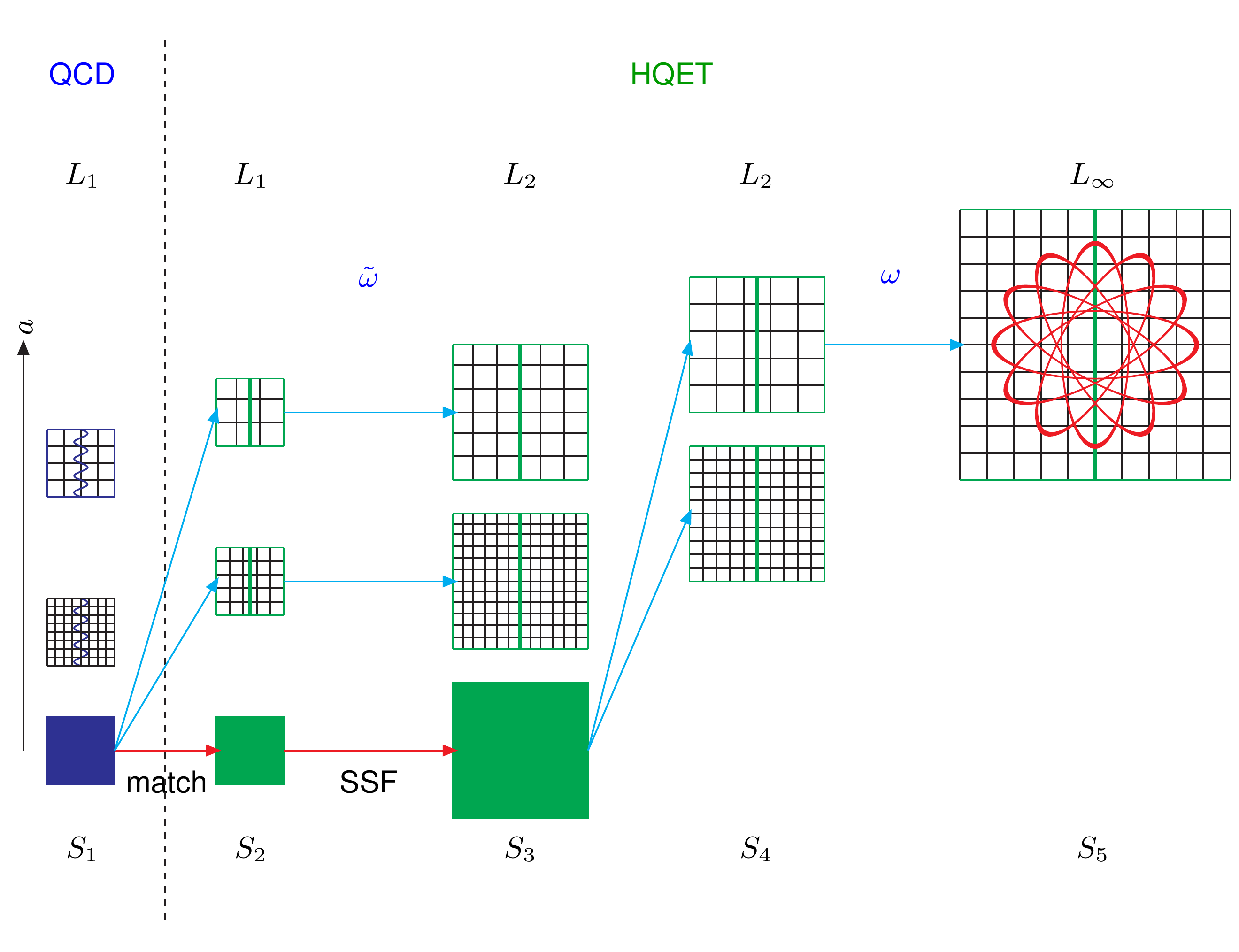}}
\end{center}
\vspace{-0.7cm}
\caption[]{\footnotesize 
Illustration of the strategy. Our numerical application uses much finer
resolutions $a/L$ than those shown here.
For each step $i$ a series of simulations $S_{i}$ is necessary, 
they are described in the text and in Table~\ref{t:simul}.
}
\label{f:hqetstrat}
\end{figure}

We collect all HQET parameters into one vector $\omega$
with components $\omega_i\,,i=1,\ldots,5$ listed explicitly
in \tab{table_hqet_param_def}. 
In static approximation the parameters are $\omega_1,\,\omega_2$. 
When $\minv$ corrections are included,
the additional parameters in the action 
are $\omega_4,\,\omega_5$; moreover the previous parameters 
change by (partially power divergent) terms of order $\minv$.
The situation with  $\omega_3$ is more intricate.
It is needed for $\rmO(a)$-improvement of the static approximation
and for genuine $\minv$-terms at order $\minv$.
We now turn to an explanation of the various steps involved
in the determination of the $\omega_i$. Our strategy 
is illustrated in \fig{f:hqetstrat}.

\subsection{Finite volume observables}
Five observables $\Phi_i$ are required to determine the HQET parameters $\omega_i$. 
We choose $\Phi_i(L,M,a)$ defined from \SF correlation functions by 
forming suitable renormalized combinations. They are universal, which means in 
particular that their continuum limit $\Phi_i(L,M,0)$ exists. 

As variables we have chosen the box size $L$ (which plays the r\^ole of a kinematical 
variable), the RGI mass of the heavy quark, $M$, and the lattice spacing $a$. Equivalent 
but dimensionless variables are a non-perturbative
running coupling $\gbar(L)$ (e.g. in the \SF scheme), the combination
$z=ML$ and the resolution $a/L$. 
The light quark masses are assumed fixed (in the numerics they are set to zero).

Most of the details of
our choice follows closely \cite{hqet:pap4}; for others we refer 
the reader to \app{a:choice}.
Here we note only a few properties. 
As $L$ becomes large, $\Phi_i\,,\;i=1,2$ 
tend to the B-meson mass and the logarithm of the
B-meson decay constant,
up to kinematical constants. They are thus 
mainly determining $\omega_{i\leq2}$. 
Correspondingly 
$\Phi_4$ and $\Phi_5$
allow easy access to the $\minv$-parameters in the Lagrangian
and $\Phi_3$ to the correction term of the current with coefficient $\cah{1}$.

By a choice of notation (e.g. considering $\ln(\ampl)$ instead of 
$\ampl$), we arranged our
finite volume observables to be linear in $\omega$,
\bes
 \Phi(L,M,a) = \phistat(L,a) + \phimat(L,a)\,\omega(M,a)\,. 
 \label{e:phiexp}
\ees

By construction $\phimat$ is a block diagonal matrix, which is explicitly given 
in terms of (bare) correlation functions in HQET. The inhomogeneous pieces $\phistat_i$ 
involve just correlation functions in the static approximation. Their continuum limits
$\phistat_i(L,0)$ exist for $i>2$, while for $i=1,2$ additive renormalizations
are necessary. They are contained in the second term of \eq{e:phiexp}.
In the case $i>2$, the two terms of the right hand side of \eq{e:phiexp}
are then computed separately, see section~\ref{s:split}.

The HQET parameters $\omega$ are defined by matching at a certain value of $L=L_1$ 
the HQET observables $\Phi_i$ to the corresponding values $\Phi^\mrm{QCD}_i$ 
computed in QCD. These QCD observables are first extrapolated 
to the continuum limit (indicated by $S_1$ in \fig{f:hqetstrat})
\bes
  \label{e:phiqcd}
 \Phi^\mrm{QCD}_i(L_1,M,0)=\lim_{a\to 0} \Phi^\mrm{QCD}_i(L_1,M,a)\,.
\ees

\subsection{Matching}

The matching scale $L_1$ must be chosen such that ${1/L_1} \ll \mbeauty$ 
to allow for a precise expansion in $\minv$. Moreover, we want lattice spacings 
of order $10^{-2}\,\fm$ in order to keep $a\mbeauty < 1/2$ while performing
the continuum extrapolation \eq{e:phiqcd}. These constraints lead to 
a box size $L_1\approx 0.4\,\fm$ \cite{hqet:pap3,hqet:pap4}.

The HQET parameters are then defined by imposing
\bes
  \label{e:matchingcond}
  \Phi(L_1,M,a) = \Phi^\mrm{QCD}(L_1,M,0) \,
\ees
for any value of the lattice spacing. By inverting 
\eq{e:phiexp} one obtains
\bes
  \label{e:match}
  \tilde{\omega}(M,a) \equiv \phimat^{-1}(L_1,a)\,
  [\Phi^\mrm{QCD}(L_1,M,0)-\phistat(L_1,a)]\,.
\ees
Through our choice the matrix $\phimat$ is of the form
\bes
  \label{e:AB}
  \phimat = \pmat{C & B \\ 0 & A}\,,\quad \phimat^{-1}= 
            \pmat{C^{-1} & -C^{-1}BA^{-1} \\ 0 & A^{-1}}\,,
\ees
where $C=\diag(L,1)$.  The $3\times3$ sub-matrix  $A$ has a further
upper triangular structure. It is written explicitly in \app{a:ssf},
together with its inverse.

\subsection{Step scaling}
For reasonable resolutions $L_1/a\geq\rmO(10)$, only
lattice spacings $a\lesssim0.05\,\fm$ are accessible,
while for standard large volume HQET computations we
would also like larger $a$. 
In order to obtain $\omega$ also there, 
we first need $\Phi(L,M,0)$ at larger $L=L_2$. We simply use
the values $\tilde{\omega}(M,a)$ of \eq{e:match} and determine 
the continuum limit of the HQET observables at $L_2$ 
(indicated by $S_3$ in \fig{f:hqetstrat})
\bes
\label{e:phiL2}
  &&\Phi(L_2,M,0) =  \lim_{a\to0}
  \left\{ \phistat(L_2,a) + \phimat(L_2,a)\,\tilde{\omega}(M,a)  \right\}\,. 
\ees 
This can be done as long as the lattice spacing is common to 
the $n_2=L_2/a$ and $n_1=L_1/a$-lattices and 
\bes
   s =L_2/L_1 =  n_2/n_1  
\ees
is kept at a fixed, small, ratio.\footnote{
The ratio $s$ is required to be fixed for
the cutoff effects to be a smooth function of $a/L_i$.  
}
We choose $s=2$ in the numerical 
application. Typical resolutions are $L_1/a=\rmO(10)$, see \sect{s:num}.
This procedure which takes us from $L$ to $sL$ with a finite
scale factor $s$ is called step scaling \cite{alpha:sigma}.
The explicit forms of the step scaling functions are given in~\app{a:ssf}.

\subsection{HQET parameters}
The parameters $\omega_i$ for use in large volume (see \sect{s:spect})
are finally obtained from
\bes
  \label{e:fin}
  \omega(M,a) \equiv \phimat^{-1}(L_2,a)\,
  [\Phi(L_2,M,0)-\phistat(L_2,a)]\,
\ees
through HQET simulations (indicated by $S_4$ in \fig{f:hqetstrat}) 
with typical resolutions $L_2/a=\rmO(10)$. 
This results in lattice spacings of $a\approx 0.1\,\fm$ and smaller, as
desired for the applications. 
While the final results $\omega(M,a)$ do depend
on all details of the regularization (gauge action etc.),
the intermediate observables $\Phi_i(L_2,M,0)$ may be used universally
(e.g. together with an implementation of \eq{e:fin} with 
a different regularization).

\subsection{Splitting lowest order and first order in $\minv$
\label{s:split}}

In \sect{s:ambig} we discussed that the splitting 
of a prediction into different orders of the 
$\minv$-expansion is not unique. Nevertheless, it is of 
interest to organize the calculation into a static one
and the remainder. In this
way one can judge on the generic size of $\minv$ corrections and 
thus get some indication of the {\em asymptotic}
convergence of the series. A second reason is that the static
theory is $\Oa$ improved when  $\omega_3$ is
included and then represents the improvement coefficient
$a\castat=\omega_3$ \cite{zastat:pap1}. 
Thus the by far dominating part of the result can be extrapolated
quadratically in $a$ and only a small correction has to be
extrapolated linearly to the continuum limit.
Since a non-perturbative determination
of $\castat$ has not been carried out, we will here use its
one-loop perturbative value
\cite{castat:filippo}. We anyway carried out quadratic extrapolations in $a$ 
and then studied the effect of
incomplete improvement. We will see in \sect{s:num} that $\castat$ is of little
relevance. Therefore we profit from the $\Oa$-improvement of the static theory, 
irrespective of the precise value of $\castat$. 

To summarize, the static approximation is defined
by using exactly the same formulae as in HQET with $\minv$ 
but setting
\bes
  \omega_3^\stat &=&  a\castat \\
  \omega_4^\stat &=& \omega_5^\stat =0\,, \quad \Phi_{i>2}^\stat =0 \,.
\ees
All quantities are determined once in this static approximation
and once with $\minv$ terms included and then the pure $\minv$-corrections
are given by
\bes
  \Phi^\first = \Phi - \Phi^\stat \,,\; \omega^\first = \omega - \omega^\stat\,.
\ees
For any new observables, in particular in large volume,
the  static results are given by inserting $\omega^\stat$ into the
expressions such as \eq{e:expandfb}
and the $\minv$-correction by inserting $\omega^\first$ instead.
Since, as we discussed earlier everything is to be linearized
in the $\omega$, the ``full'' result up to $\minv^2$ is obtained
from summing static and $\minv$-correction or by using directly
$\omega$. We finally note that even though the ``full''
result contains $\rmO(a)$ discretization errors, it appears 
justified to extrapolate numerical data with a leading correction term $\propto a^2$
since the linear terms are suppressed by a small $\minv$ factor,
which one would estimate e.g. as $1/(\mbeauty r_0)\approx 1/10$.

%% file: tables/hqet_param_def.tex
 \begin{table}[!htb] 
 \hspace{-1.cm} 
 \begin{center} 
 \begin{tabular}{c l c c c c}
 \hline 
 $\omega_i$  & definition & & classical &&  static \\[-0.8ex] 
             &      & & value & & value\\[0.2ex]
 \hline 
 $\omega_1$  & $\mhbare$  & &  $\mbeauty$ & & $ \mhbare^\stat$  \\ [+0.7ex]
 $\omega_2$  & $\lnzahqet  $  &  &  0 & & $ \ln(\zastatrgi \Cps) $  \\ [+0.7ex] 
 $\omega_3$  & $\cahqet    $  &  &  $-1/(2\mbeauty)$ & & $ a\castat$  \\ [+0.7ex]
 $\omega_4$  & $\omegakin  $  &  &  $1/(2\mbeauty)$ & & $0$  \\ [+0.7ex]
 $\omega_5$  & $\omegaspin $  &  &  $1/(2\mbeauty)$ & & $0$  \\ [+0.7ex]
 \hline 
 \end{tabular} 
 \end{center} 
 \caption[ ]{\footnotesize Notation for HQET parameters, their values in classical 
   and static approximation.}
 \label{table_hqet_param_def} 
 \end{table}

%% file: s4.tex
\section{Numerical results \label{s:num}}

The numerical computations have been performed 
in the quenched approximation.
For the QCD part we used non-perturbatively
O($a$)-improved Wilson fermions \cite{impr:pap3},
see \cite{hqet:pap4} for details of the 
\SF implementation.
Two discretizations of the static quarks
have been considered:
the so-called HYP1 and HYP2 actions~\cite{HYP,stat:actpaper}.
In order to reduce discretization effects, 
we have implemented tree level improvement
of our QCD observables and of the HQET step scaling 
functions (see \app{a:tli}).

Our observables depend on three periodicity angles
$\theta_0,\theta_1,\theta_2$, (see \app{a:choice}). 
As in \cite{hqet:pap4} we considered all
combinations of those with
$\theta_i\in\{0,\,0.5,\,1.0\}\,,$ $ \theta_1 < \theta_2$, i.e. nine different
matching conditions. For our discussion of the 
results and the extraction of the final
HQET parameters we chose a ``standard set''
\bes
  \label{e:defaulttheta}
  \theta_0=0.5\,,\;\theta_1=0.5\,,\; \theta_2=1.0 \,
\ees 
because this yields overall the smallest statistical 
errors in the HQET parameters. We will comment on the spread
of results with other choices for $\theta_i$ as we go along.

We summarise the different simulations needed in our strategy in \tab{t:simul}.
\input{tables/simul.tex}
Since a size $L=L_1$ is used for the matching (\eq{e:phiqcd}, \eq{e:match}) 
both QCD ($S_1$) and HQET ($S_2$) are 
simulated in that volume.
The volume of space extent $L_2$ was simulated (in HQET), 
with two different sets of lattice spacings:
\begin{itemize}
\item
First ($S_3$) with the same set of lattice spacing as used in $S_2$,
to evaluate \eq{e:phiL2}.
\item
Then ($S_4$) with lattice spacing of order $0.05 \ldots 0.1\,\fm$, 
in order to compute $\omega(M,a)$ given by \eq{e:fin}. 
\end{itemize}
The choice of the simulation parameters is described in
detail in \cite{hqet:pap4}; Table~3 of that paper lists those
for $S_1$, Table~A.1 of \cite{hqet:pap1} the parameters of $S_2$ and $S_3$ and
Table~6 of \cite{stat:actpaper} the parameters of $S_4$. 
Here we just note that $L_2=2L_1$ and
$L_1$ is fixed by the Schr\"odinger functional 
coupling $\bar g^2(L_1)=u_1=3.48$. Since this condition was implemented
only within a certain precision, a small mismatch of 
$\gbar^2(L)=\tilde u_1$ used in $S_1$ and $\gbar^2(L)=u_1$ used everywhere else
has to be taken into account.
This is done in complete analogy to appendix D of \cite{hqet:pap4}. Without 
discussing the details we will quote the small corrections below.

\subsection{Observables in $L=L_1$.}

The considered observables naturally split into three classes
according to their magnitude in the $\minv$ expansion, or more precisely
their order in $z^{-1}=(ML)^{-1}$. We have
$\Phi_1=\rmO(z), \Phi_{2\ldots4} = \rmO(1)$ and
$\Phi_{5}=\rmO(1/z)$. As an illustration of the numerical
determinations, in particular the continuum limit,
we discuss examples of all three cases.
The precise definitions of $\Phi_i$ are found in 
appendix~\ref{a:choice}.

\begin{figure}[!htb]
\begin{center}
\includegraphics[width=7cm]{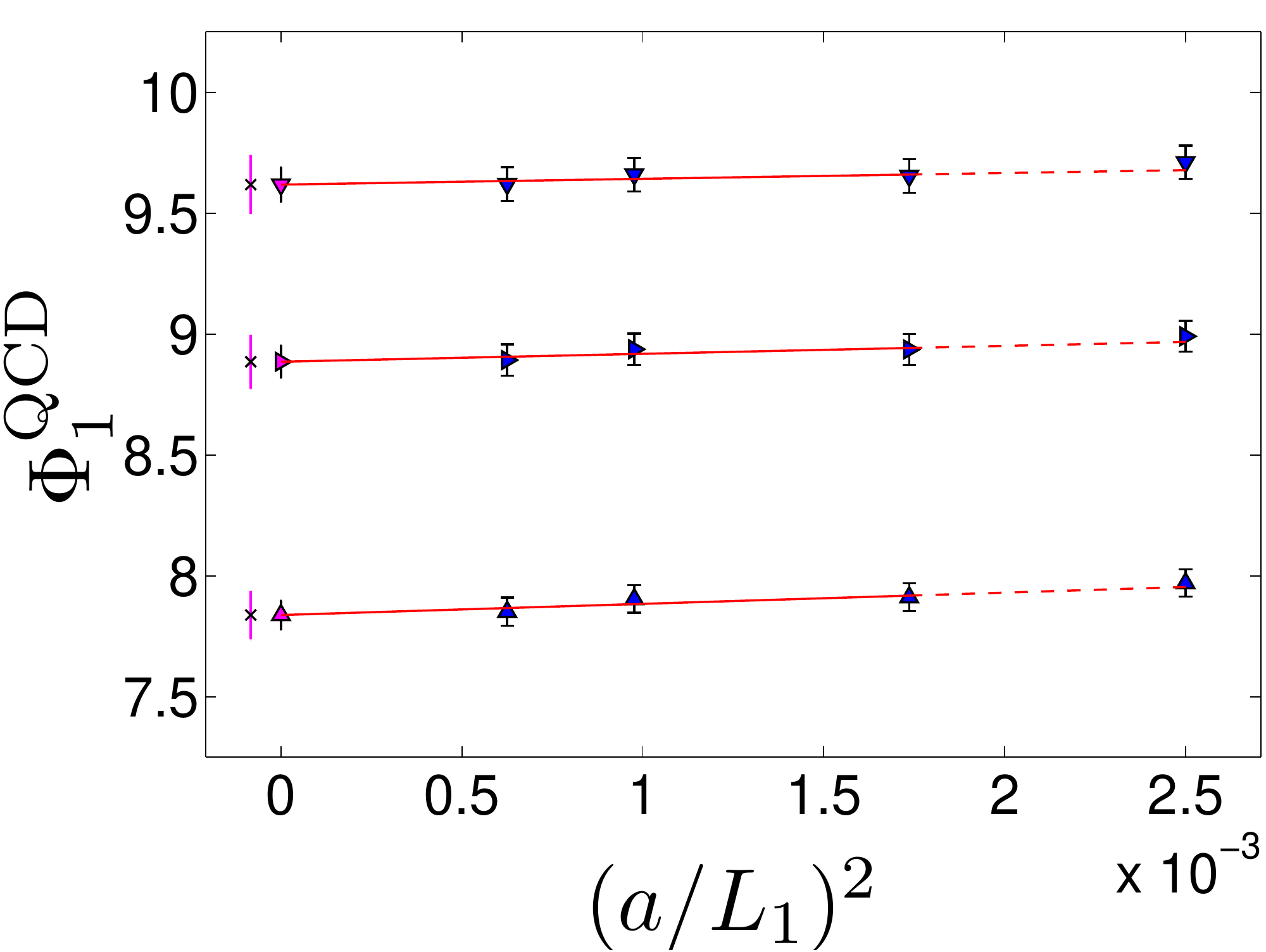}
\includegraphics[width=7cm]{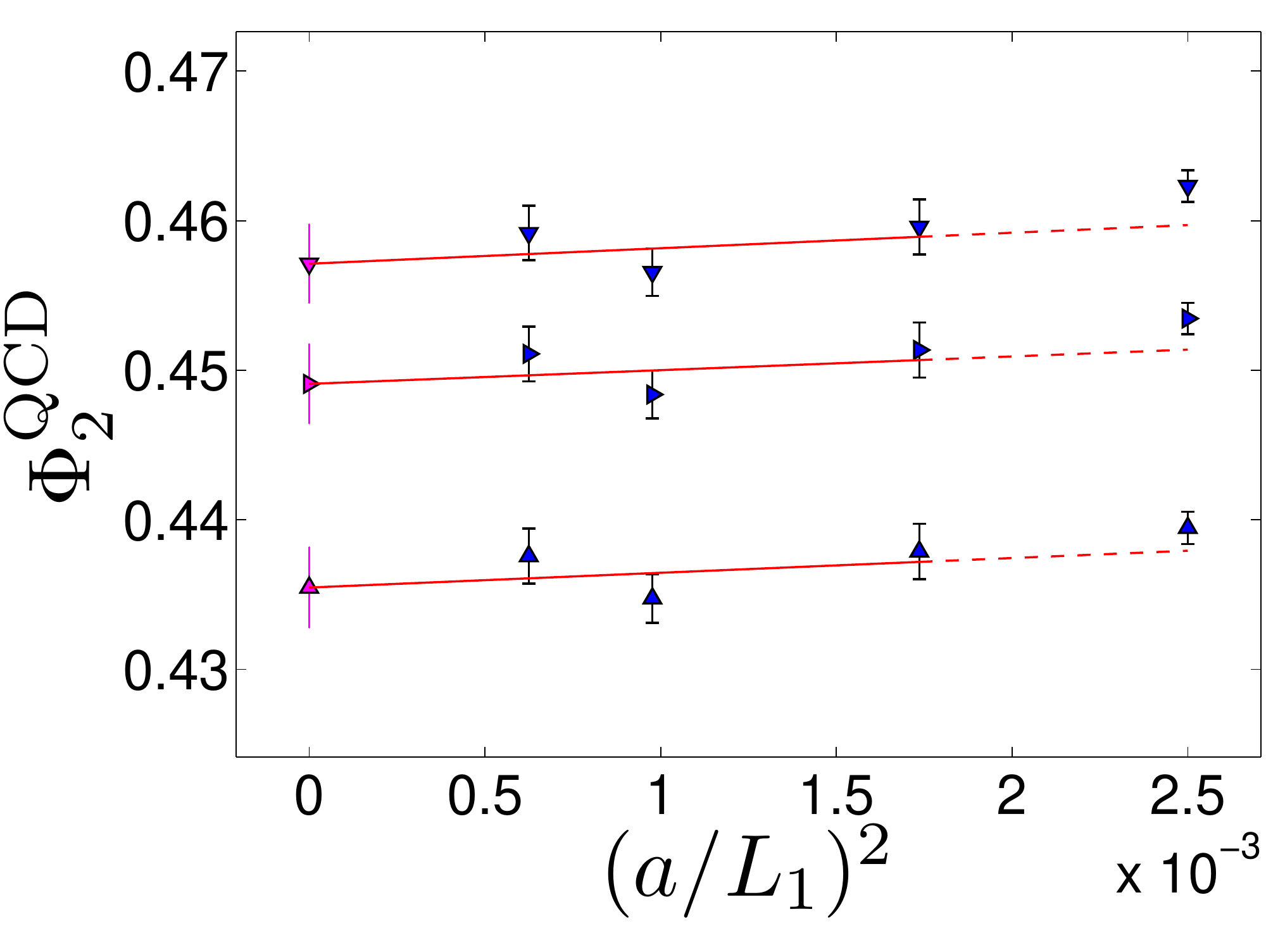}
\end{center}
\vspace{-0.7cm}
\caption[]{\footnotesize Continuum extrapolation of $\Phiqcd_1(L_1,M,a)$ and 
$\Phiqcd_2(L_1,M,a)$,
for the three different quark masses (from the lightest to the heaviest:
upward-pointing, right-pointing and downward-pointing triangle). 
As explained in the text, the final errors for $\Phiqcd_1(L_1,M,0)$ 
are shown on the very left.
Only the data in the range of the full lines are used in the 
continuum extrapolations. 
}
\label{fig:Phi1}
\end{figure}

We recall that $\Phi_1$ is simply a finite volume 
pseudo-scalar meson mass (up to a normalization) and 
determines the quark mass. 
In \fig{fig:Phi1}, we show its continuum extrapolation
for our chosen values of the heavy quark
mass, corresponding to $z = L_1M = 10.4, 12.1, 13.3$. 
We have taken into account the errors coming from the relation between
the bare quark mass and the RGI quark mass. In particular, a part 
of this error is common to the data at all lattice spacings. 
It is included after the continuum extrapolation; the final error bar
is shown on the left side of this plot (see also \cite{hqet:pap4}). 
Since $\rmO(a)$ improvement is implemented, this as well as
all other extrapolations of \eq{e:phiqcd} are done with the form 
$\Phi_i(L_1,M,a)=\Phi_i(L_1,M,0) + c_i(M)\, (a/L_1)^2$.

Out of the $\rmO(z^0)$ observables, $\Phi_2$ determines
the normalization of the axial current.  Its continuum
extrapolation, \eq{e:phiqcd}, is also illustrated in \fig{fig:Phi1}.

\begin{figure}[!htbp]
\begin{center}
\begin{tabular}{cc}
\hspace{-0.5cm}
\includegraphics[width=7cm]{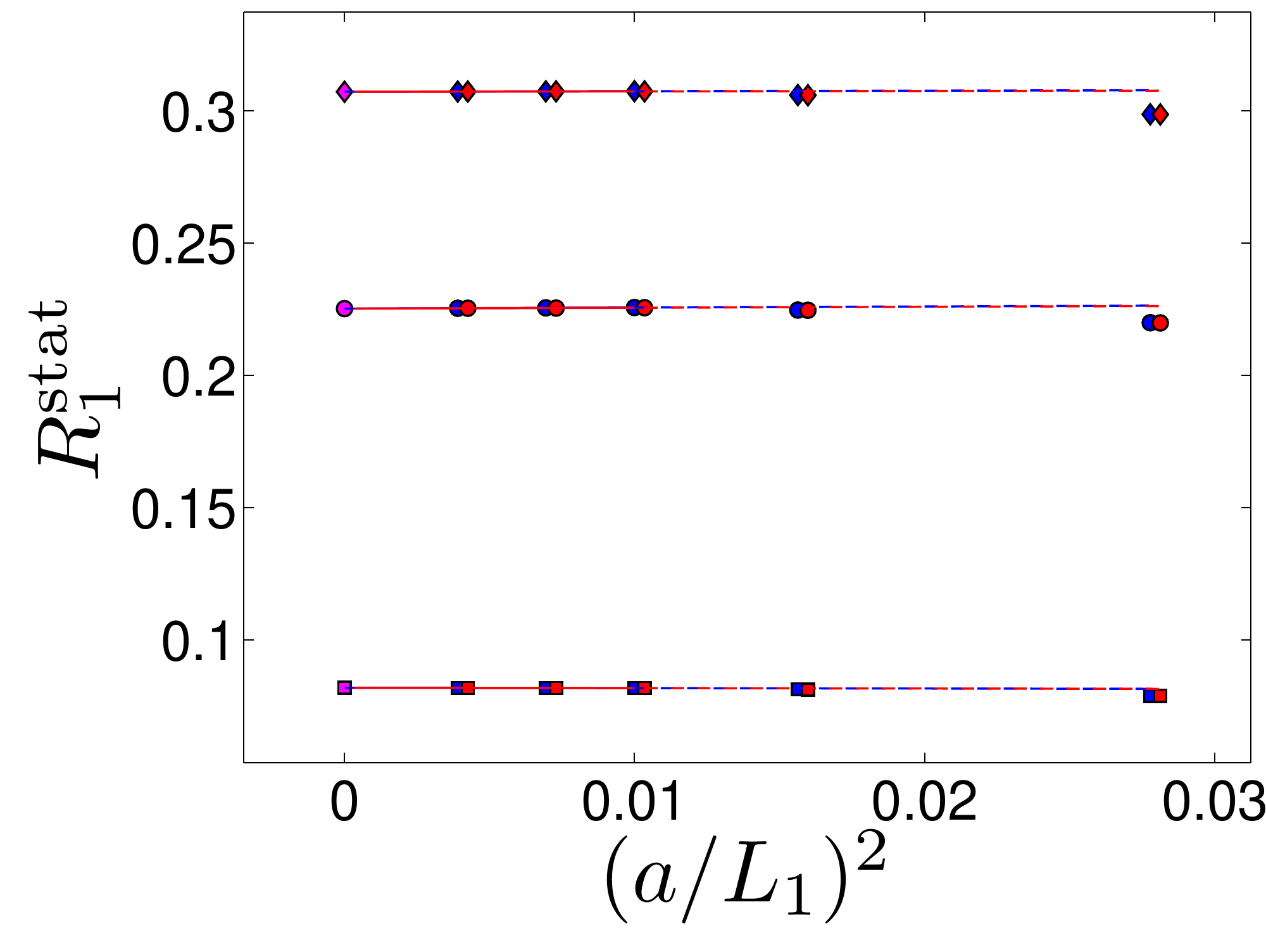}&
\includegraphics[width=7cm]{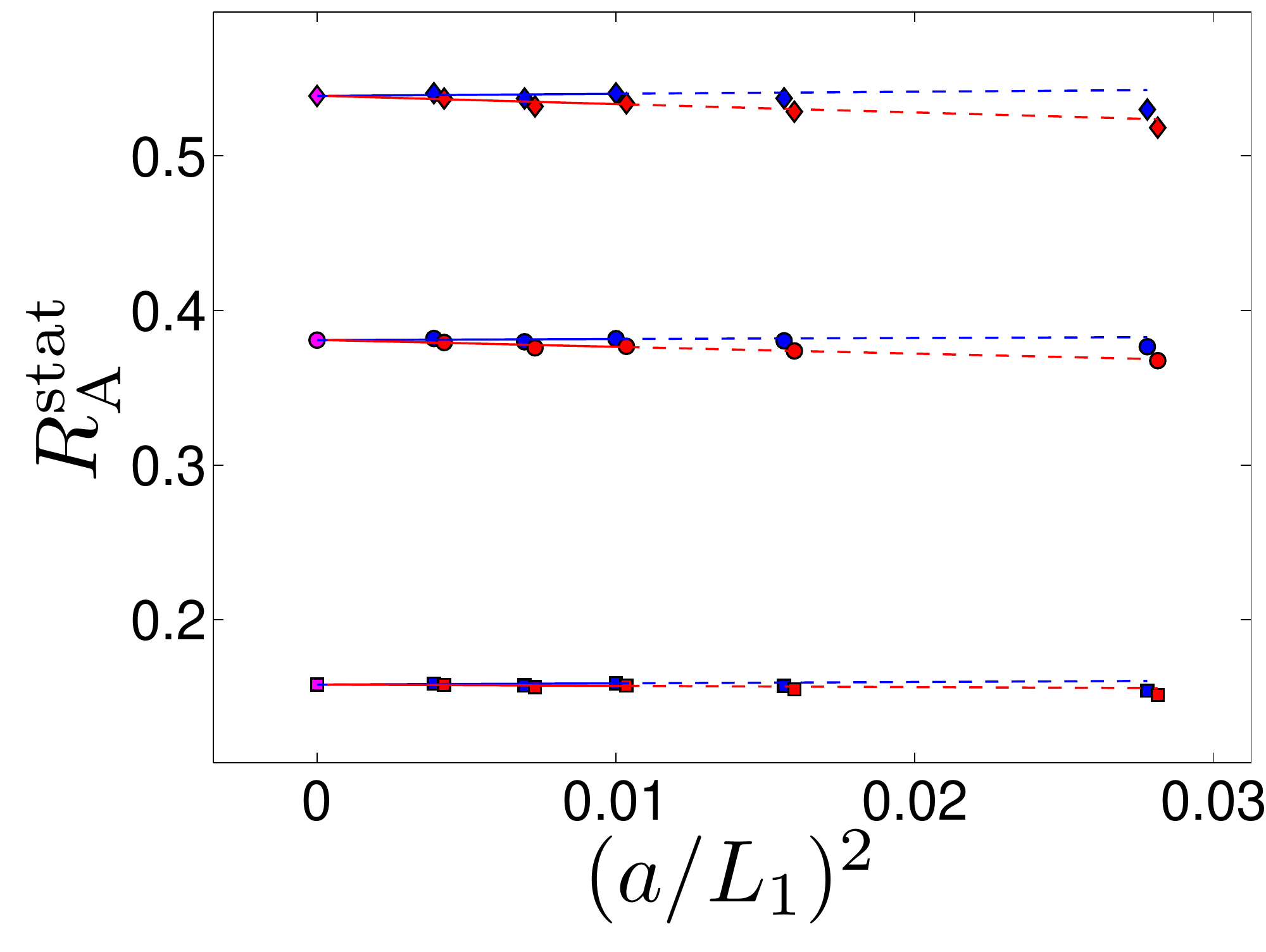}\\
\end{tabular}
\end{center}
\vspace{-0.7cm}
\caption[]{\footnotesize Continuum extrapolation of 
$\phistat_3(L_1,a)=\rastat(L_1,a)$,
and $\phistat_4(L_1,a)=\ronestat(L_1,a)$. 
The three different choices 
$(\theta_1, \theta_2) = (0, 0.5),\, (0.5, 1),\, (0, 1.0) $
are represented by squares, circles, and diamonds respectively.
In each case two different static actions, HYP1 (the red points, 
slightly shifted to the right for visibility) and HYP2 (the blue points)
are used.
The continuum limit is obtained by a
constrained fit $\phistat_i(L_1,a) = \phistat_i(L_1,0)+c_{i,j}\, a^2/L_1^2$,
with $j=1,2$ for the two different actions.}
\label{fig:etab}
\end{figure}

Some of the static quantities $\phistat_i(L_1,a)$ 
have a well defined continuum limit 
(see appendix~\ref{a:ssf}). In these cases, $\phistat_i(L_1,a)$
is replaced by $\phistat_i(L_1,0)$ in 
\eq{e:match}. We show examples of the determination
of the continuum limit $\phistat_3(L_1,0)$
and $\phistat_4(L_1,0)$ in \fig{fig:etab}. The graphs use the one-loop
values of the improvement coefficient\cite{castat:filippo}
\bes 
  \castat &=& 0.0029 \, g_0^2 \;\;\mbox{for the HYP1 action}\,.
      \nonumber\\[-1ex] \label{e:castat1} \\[-1ex]   \nonumber
  \castat &=& 0.0518 \, g_0^2 \;\;\mbox{for the HYP2 action}\,,
\ees
As a test in how far the remaining uncertainty in $\castat$ matters,
we repeated the calculation setting $\castat=0$ instead. The change
is invisible in the graphs and in general, for all the
continuum extrapolations of this paper, the changes in continuum values
are much smaller than our quoted errors. We thus consider \eq{e:castat1}
entirely sufficient. 

\begin{figure}[!htbp]
\begin{center}
\begin{tabular}{cc}
\hspace{-0.5cm}
\includegraphics[width=7cm]{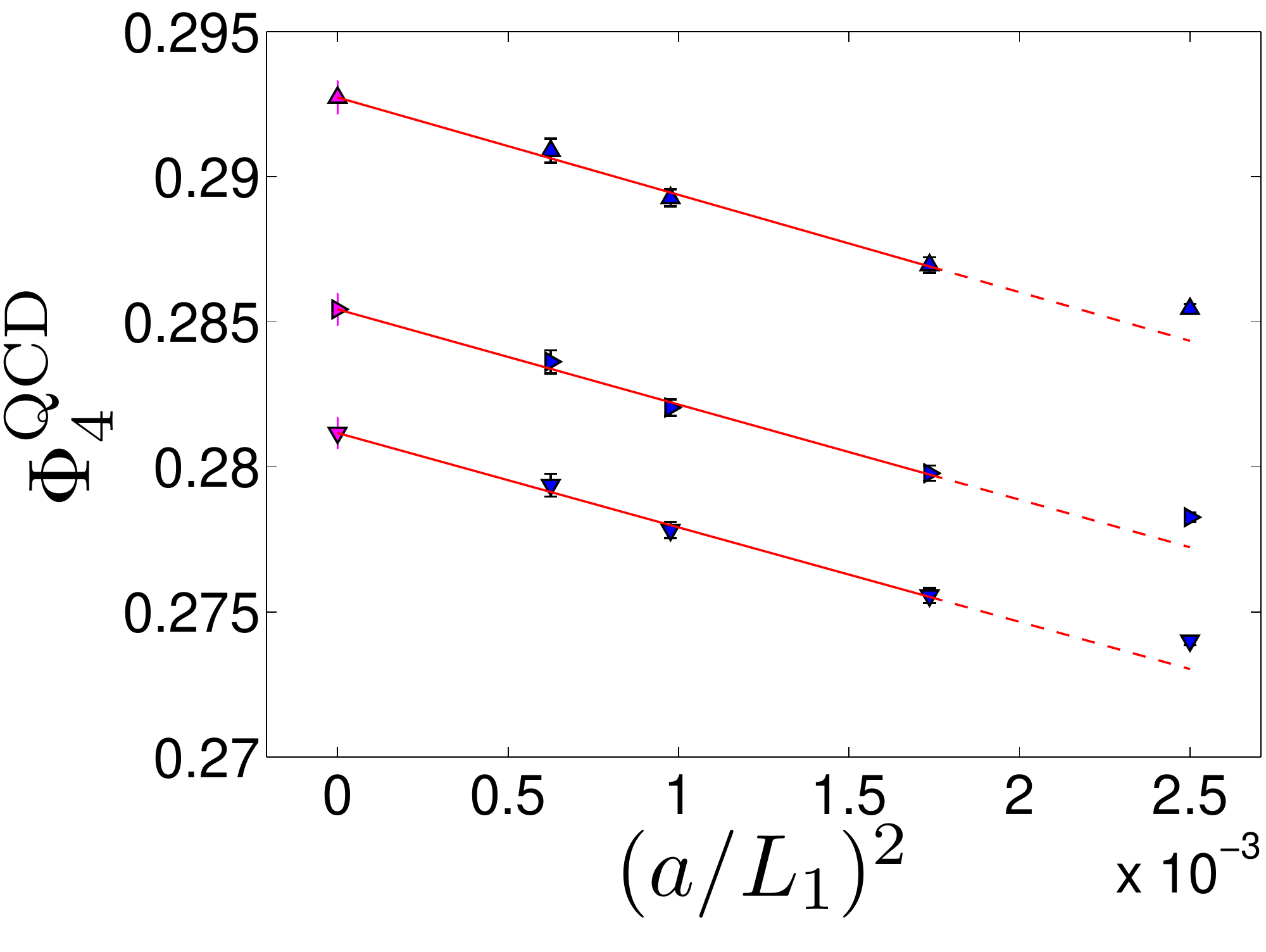}&
\includegraphics[width=7cm]{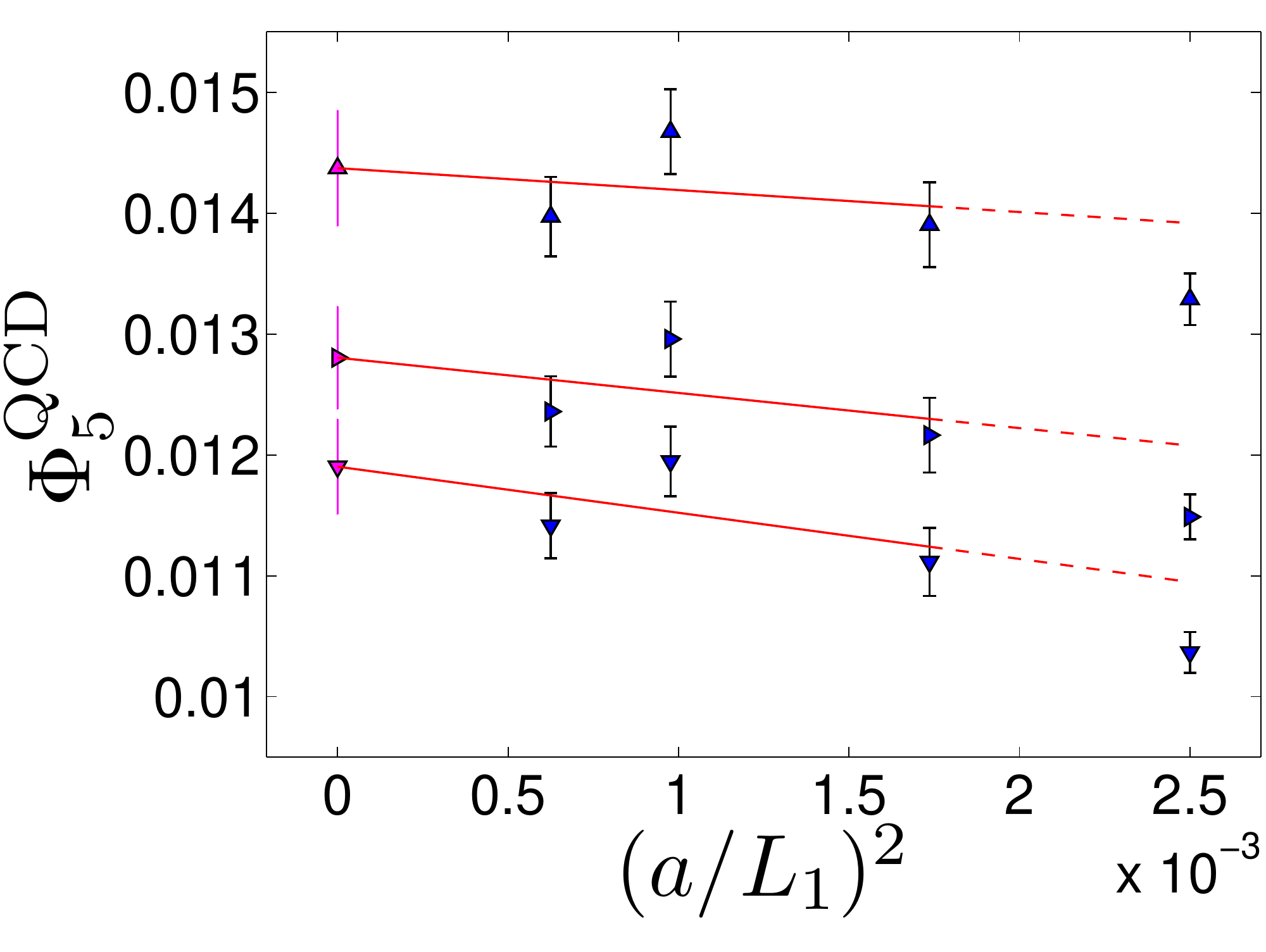}\\
\end{tabular}
\end{center}
\vspace{-0.7cm}
\caption[]{\footnotesize Continuum extrapolation of the QCD observables
$\Phi^\mrm{QCD}_4(L_1,M,a)$ and $\Phi^\mrm{QCD}_5(L_1,M,a)$ for the standard 
$\theta$'s. The three different quark masses are shown, with the same 
conventions as in \fig{fig:Phi1}.
}
\label{fig:phi45}
\end{figure}

We turn now to $\Phi_4$ and $\Phi_5$, needed for the determination
of the $1/\mbeauty$ parameters $\omegakin$ and $\omegaspin$.
Their continuum extrapolations are shown in \fig{fig:phi45}.
Due to the exact spin symmetry of the static effective theory,
$\Phi_5$ has no static contribution. 
Note that both $\Phi_5$ and 
the pure $\minv$ part of $\Phi_4$, obtained
after subtraction of $\phistat_4(L_1,0)$ (see \fig{fig:etab}), 
are an order of magnitude smaller than $\Phi_4$. 
As expected $\Phi_4-\phistat_4$ and $\Phi_5$ are 
decreasing functions of the quark mass.

\subsection{Observables for $L=L_2$.}

The step to observables in the larger volume is described
by \eq{e:phiL2}. It
can be broken up into several step scaling functions
defined in appendix~\ref{a:ssf}, which individually 
have a continuum limit. Discussing them one by one 
would be too lengthy and is also not very illuminating. 
We follow exactly \sect{s:param} and insert \eq{e:match} into \eq{e:phiL2} 
lattice spacing by lattice spacing and then extrapolate 
$a/L_2\to0$. 

\begin{figure}[!htb]
\begin{center}
\begin{tabular}{cc}
\hspace{-0.6cm}
\includegraphics[width=7cm]{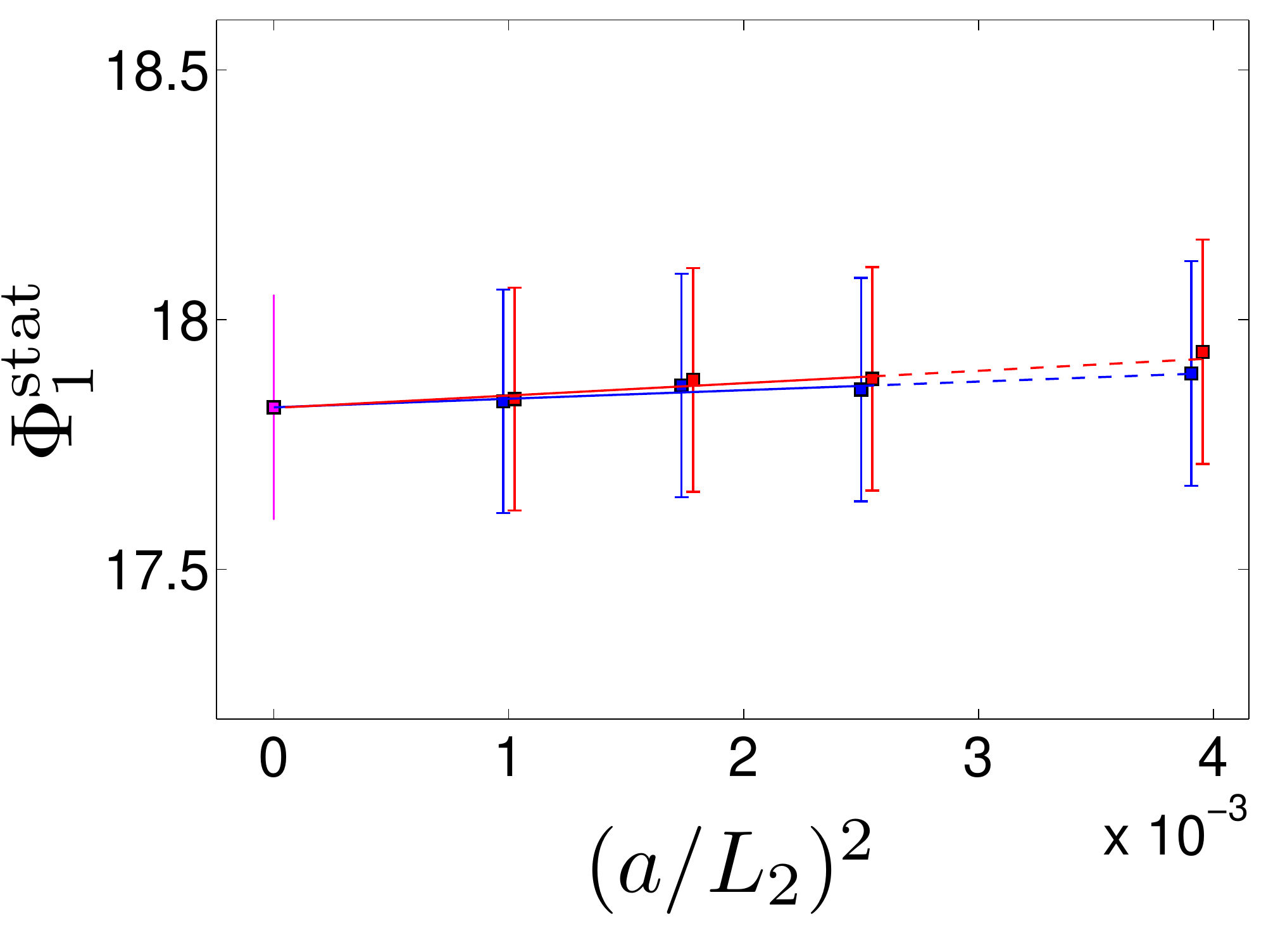}&
\includegraphics[width=7cm]{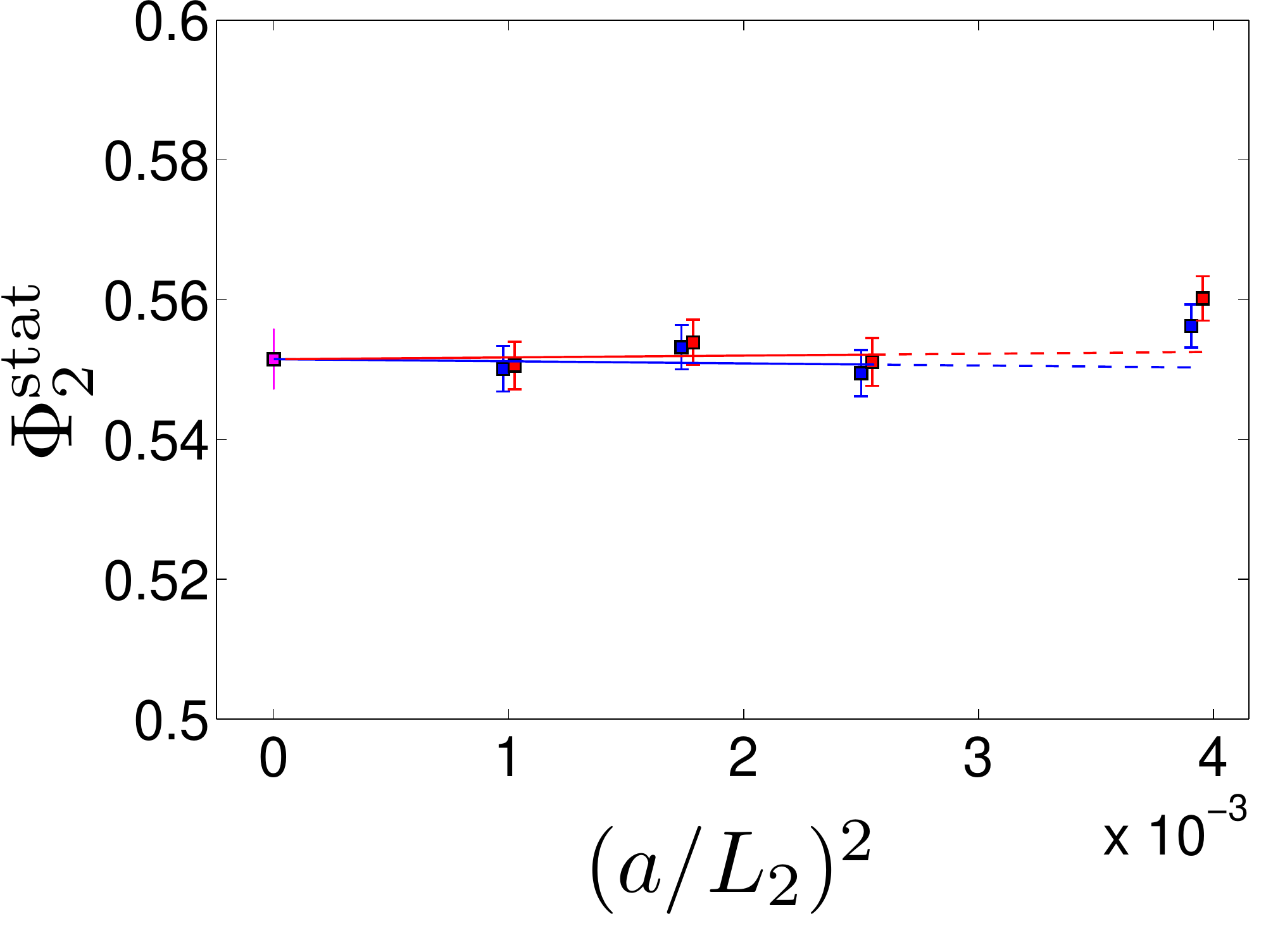}\\
\includegraphics[width=7cm]{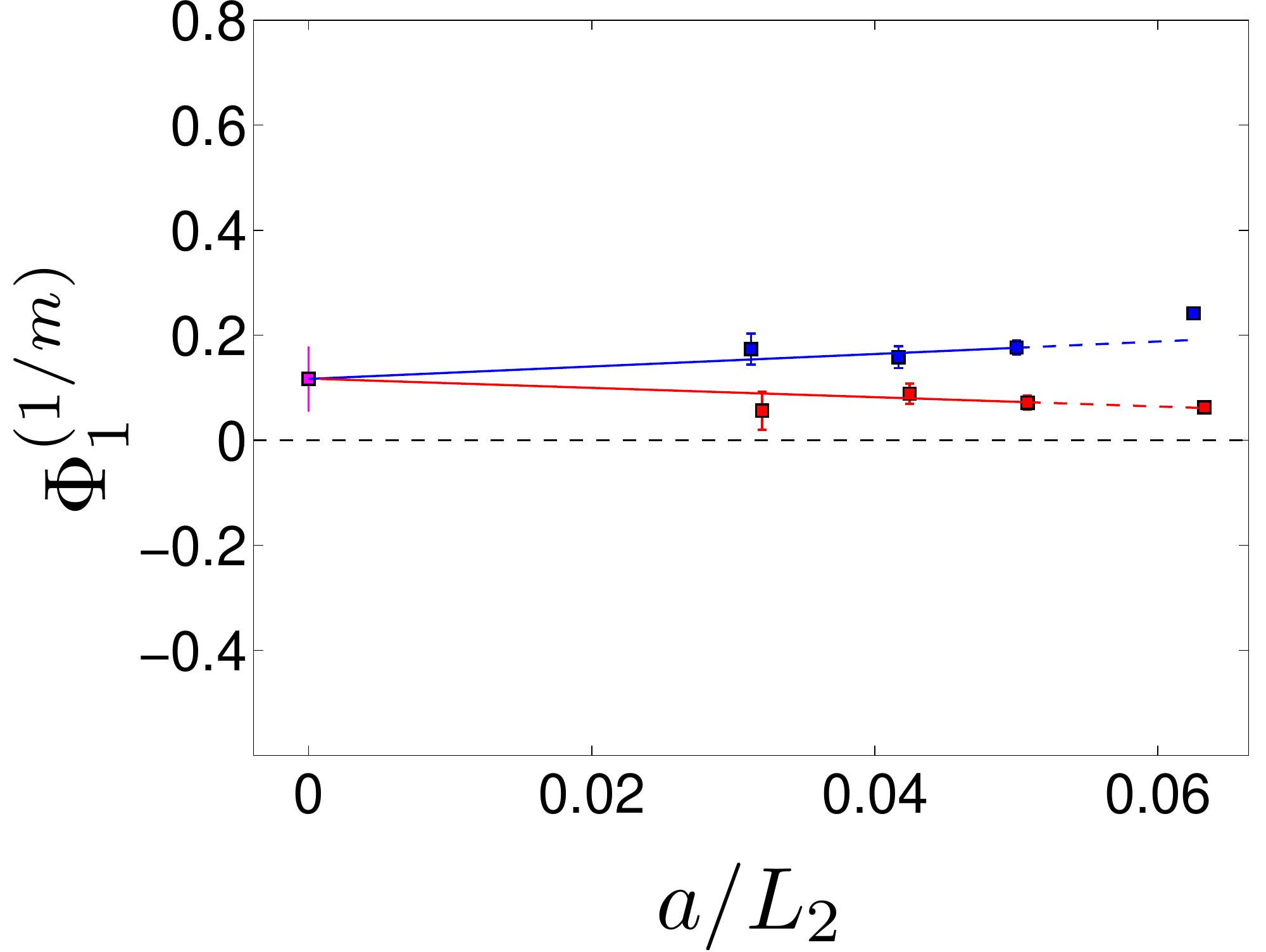}&
\includegraphics[width=7cm]{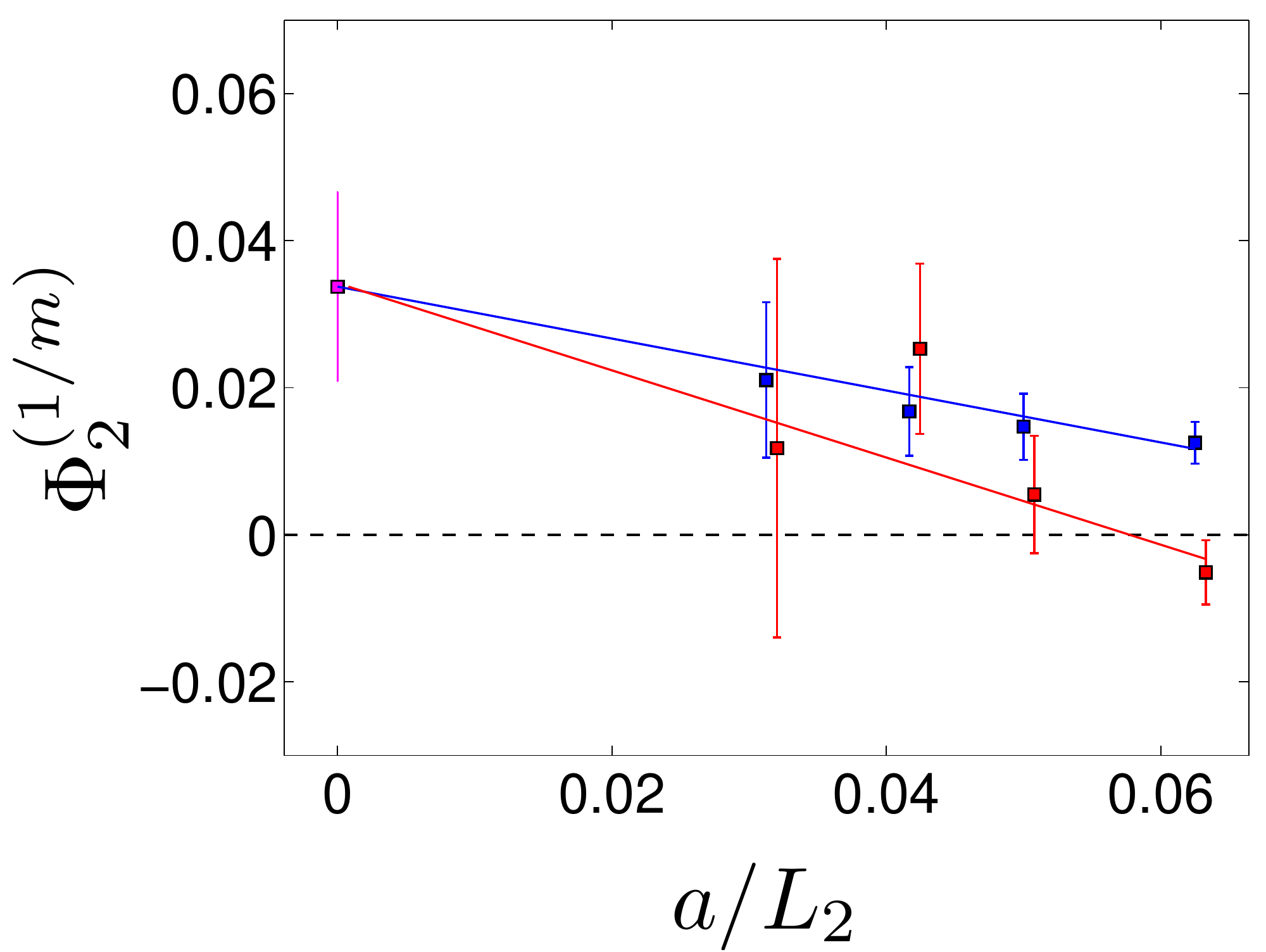}\\
\end{tabular}
\end{center}
\vspace{-0.7cm}
\caption[]{\footnotesize Continuum extrapolation of 
$\Phi_1(L_2)$ and $\Phi_2(L_2)$
for the central mass $z=L_1 M = 12.1$
and for the standard $\theta$'s.  The scale of the y-axis 
of static (top) and $\first$ parts is chosen to be equal.
As explained in the text, the static parts are extrapolated
quadratically in $a$ while the $\first$ contributions 
are extrapolated linearly in $a$.
The conventions are the same as in \fig{fig:etab}.}
\label{fig:phi12_L2}
\end{figure}

Two examples are shown in \fig{fig:phi12_L2}.
We observe that having the data for two different static actions is very useful to
constrain the continuum limit, particularly so for the $\minv$ parts.
Also the resolution $L_2/a=32$ which is
in addition to those of \cite{hqet:pap4} helps a lot 
(the reader is invited
to compare $\Phi_1^\first$ in \fig{fig:phi12_L2}
to Fig.~5 of \cite{hqet:pap4}).
In all cases the $\minv$ corrections
are much smaller than the leading terms, suggesting a good {\em asymptotic}
convergence of the $\minv$ expansion.
The precision of the $\minv$ term is worse than the
static one for $\Phi_2$, partly since the latter can be extrapolated
quadratically in $a$ to the continuum. For $\Phi_1$ this is different
because the overall error contains a large piece coming from
the renormalization factor determining the RGI quark mass in QCD. 

\begin{figure}[!htb]
\begin{center}
\hspace{-0.5cm}
\includegraphics[width=7cm]{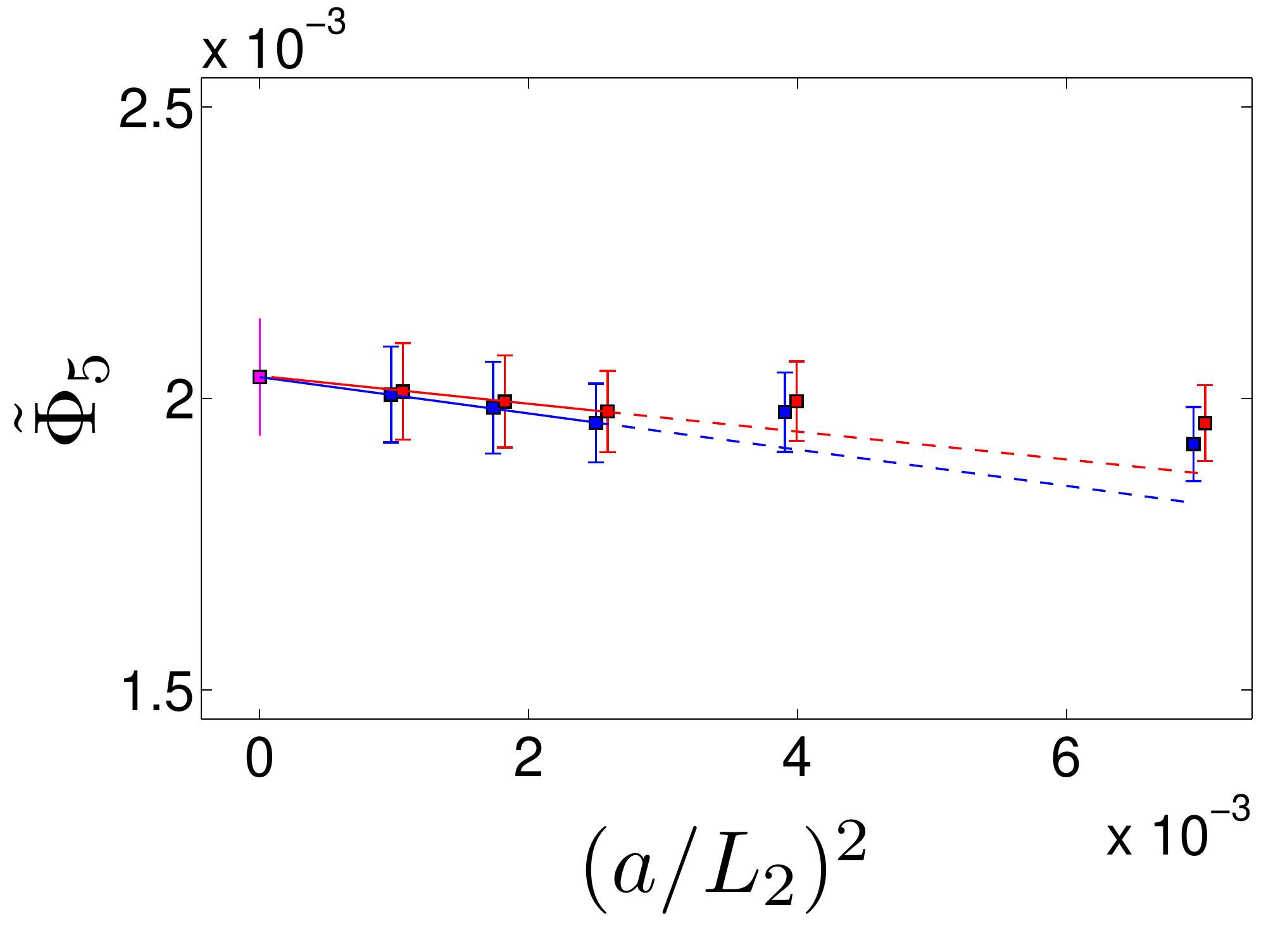}
\end{center}
\vspace{-0.7cm}
\caption[]{\footnotesize 
Continuum extrapolation of $\tilde \Phi_5(L_2)$
for our standard set of $\theta_i$. The central quark 
mass is shown. The conventions are the same as in \fig{fig:etab}.
}
\label{fig:phi5L2}
\end{figure}

For reasons of numerical precision, $\Phi_5$ (and only
$\Phi_5$) is not 
computed exactly as described in \sect{s:param}. Its 
definition (\app{a:choice}) involves the propagation of a heavy quark 
over a distance $T=L$, introducing significant statistical
errors for large $L/a$ in the effective theory. These become unpleasantly large 
in \eq{e:phiL2}, more precisely in $\phimat_{55}(L_2,a)$.  
We therefore replace $\Phi_5(L_2)$ by $\tilde\Phi_5(L_2)$ 
differing only by the choice $T=L/2$. An obvious question is why this
is not done already for $L=L_1$. The reason is that $\tilde\Phi_5(L_1)$
turns out to be quite a bit smaller than its natural order
of magnitude of $\rmO(1/z)$. In such a situation $\rmO(1/z^2)$
terms may be numerically comparable and the $\minv$ expansion may be
compromised in the matching step. We therefore chose the described 
solution, even if it is lacking elegance.~\footnote{
We have of course also determined $\tilde\omega_5$ from $\tilde\Phi_5(L_1)$.
Despite the potentially large $1/m^2$ contributions, the result for
 $\tilde\omega_5$ changes only by about 15\%. 
} 
The continuum extrapolation of $\tilde\Phi_5(L_2)$ is shown 
in \fig{fig:phi5L2}.

Since the computed $\Phi(L_2,M,0)$ may be used in the
subsequent step \eq{e:fin} also with lattice discretizations which 
differ from ours we list $\Phi_i(L_2,M,0)$ in~\tab{t:phiL2stat}
and~\tab{t:phiL2onem}. Starting from these 
numbers, the remaining computations to obtain the HQET parameters for a different 
lattice action require a very modest effort. 
\input{tables/phiL2stat.tex}

\input{tables/phiL2onem.tex}

\subsection{HQET parameters}

\subsubsection{Renormalization group invariant b-quark mass}

For each value of $z=ML_1$, \eq{e:fin} yields
the desired HQET parameters. However, for future use we want
to list them for $M=\Mbeauty$. This saves space here and
somebody doing a computation in the future does not have to
fix the right quark mass. The b-quark mass was already obtained
in \cite{hqet:pap4} from the experimental spin-averaged
$B_s$ meson mass and $r_0=0.5\,\fm$. 
Here we repeat its determination using the mass of the
pseudoscalar $B_{\rm s}$ meson as experimental input. 
This is the natural quantity since we also start from the effective mass 
defined from the $f_{\rm A}$ correlator to fix the b-quark mass in the small volume. 
Moreover, we have improved some
of the necessary steps through the finer resolution
$a/L_2=1/32$, and the use of tree level improvement (see~\app{a:tli}). 
With our new, improved determinations \cite{lat09:tereza} 
using the GEVP method~\cite{gevp:pap} we found~\footnote{
The reader might wonder why for $r_0 \Mbeauty$ we don't obtain 
a better precision compared to the result quoted in 
\cite{hqet:pap4}. The reason is that the uncertainty which affects 
the b-quark mass is dominated by the renormalization of the quark
mass in QCD\cite{hqet:pap4}.}
\bes
  r_0 \Mbeauty^\stat =  17.12 (26)\,,\; 
  r_0\Mbeauty^\mrm{HQET} =  17.38(28) \;.
  \label{e:mbn2} 
\ees
These results are in perfect agreement with the ones obtained 
when using the large volume numbers $\Ekin,\Estat$ of \cite{hqet:pap4}.
In the following we use \eq{e:mbn2} and the knowledge 
of $L_1/r_0$ and interpolate all results
quadratically in $z$ to $z_\beauty = 12.30(19)$ in the static 
approximation and to $z_\beauty =  12.48(20)$ at first order in $\minv$.

\subsubsection{Bare parameters $\omega_i$}

Our $\omega_i$ determined from \eq{e:fin} and simulations $S_4$
are listed
in \tab{table_hqet_param} for our standard $\theta-$combinations. 
The errors take all sources into account
(through a jackknife analysis incorporating all steps),
but one has to be aware that there are very significant 
correlations in the parameters. We discuss these correlations
in \app{a:cov} and provide them in tables available on the Internet.
In static approximation a small shift proportional to
$\tilde u_1 - u_1$ (see the discussion at the beginning of this
section) is applied. 
We refer to Appendix D of~\cite{hqet:pap4} for the case of the b-quark mass, 
and with similar considerations we 
have evaluated the effect on the current renormalization, $\lnzastat$.
We found~\footnote{These shifts have to be added
to the raw numbers computed with our simulations $S_1,\ldots,S_4$
but are already included in the numbers given in~\eq{e:mbn2} 
and in \tab{table_hqet_param}.}
\bes
\theta_0=0 \,,\,\, &\qquad   r_0 \Delta \Mbeauty  = -0.042(20)\,, &
\qquad \Delta\lnzastat = 0.007(5)
\,,\\
\theta_0=1/2\,, &\qquad   r_0 \Delta \Mbeauty  =  0.009(11)\,, &
\qquad \Delta\lnzastat = 0.008(5)
\,,\\
\theta_0=1 \,,\,\,  &\qquad   r_0 \Delta \Mbeauty  =  0.150(45)\,, &
\qquad \Delta\lnzastat = 0.010(4) 
\;.
\ees

\input{tables/hqet_param_interp.tex}

Generically the bare parameters $\omega_i$ are completely non-universal
and depend on all details of the action. However here we are working in the quenched
approximation. In this situation the HQET action can in principle be determined
independently of the light quarks. Thus $\omega_1,\omega_4$ and $\omega_5$ 
can be used for any light quark action also
different from our specific one. 

\begin{figure}[!htb]
\begin{center}
\begin{tabular}{cc}
\hspace{-0.5cm}
\includegraphics[width=7cm]{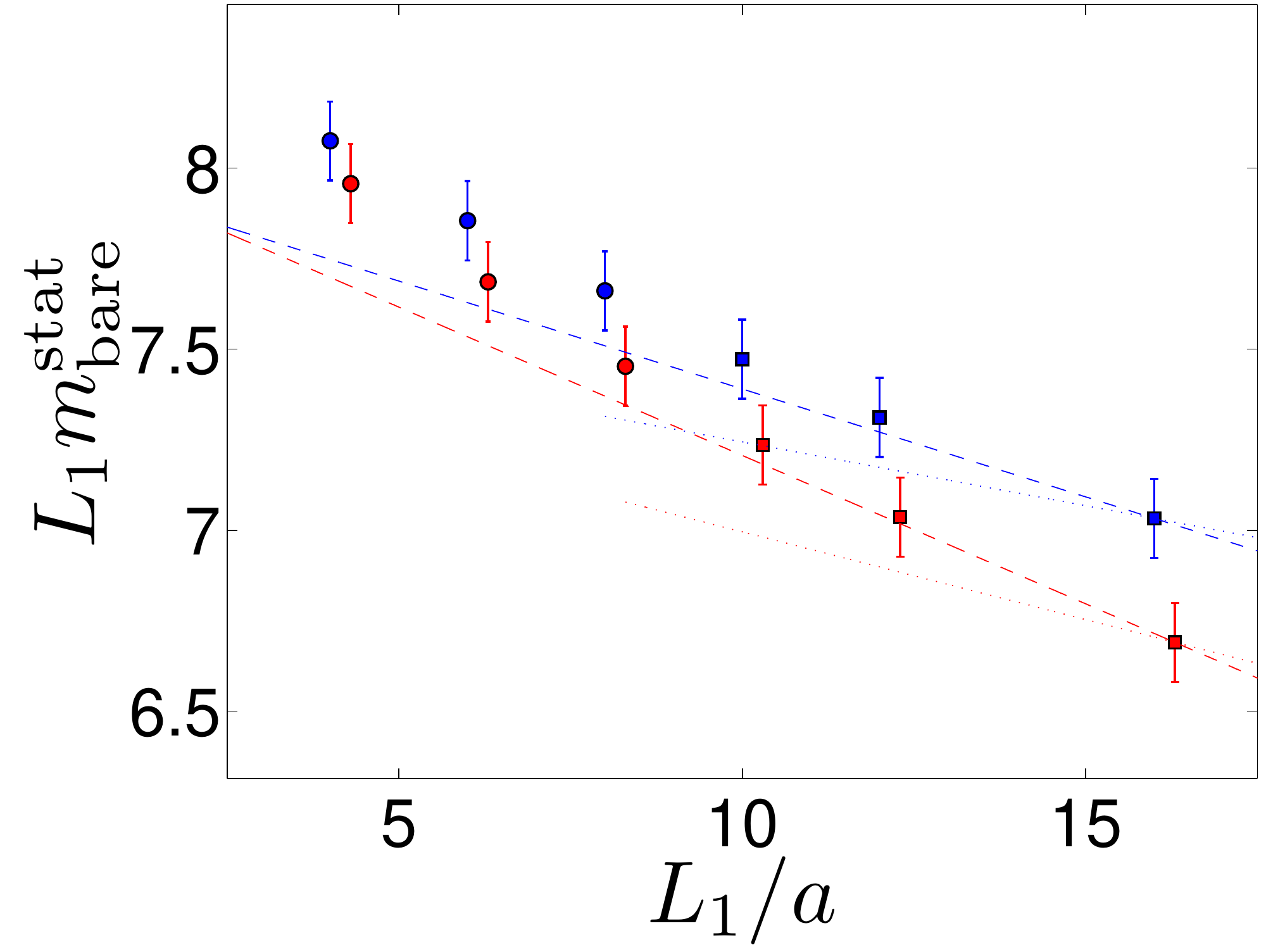}&
\includegraphics[width=7cm]{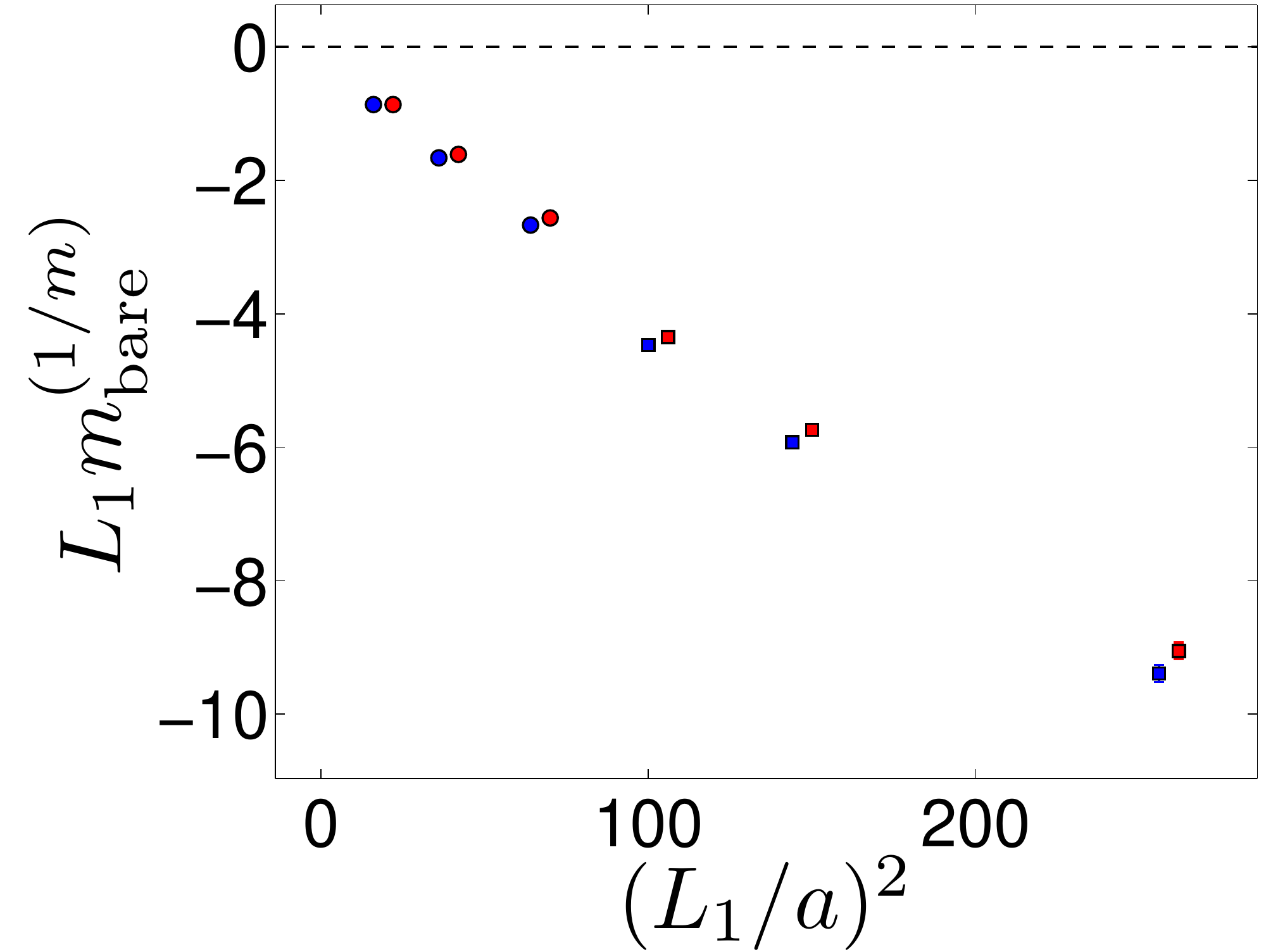}
\end{tabular}
\end{center}
\vspace{-0.7cm}
\caption[]{\footnotesize 
Static contribution (left) and $1/\mbeauty$ correction (right)
to the heavy bare quark mass. 
In the static case, we also show a comparison with
standard and boosted perturbation theory~\cite{stat:actpaper}, as described
in the text.
Data for $\omega_1$ from \tab{table_hqet_param} (large volume simulations $S_4$) 
are represented by circles (the three points on the left), while 
$\tilde{\omega}_1$ obtained from the small volume simulations $S_2$ 
are represented by squares. We use the colour blue for HYP2 
and red for HYP1 (also slightly shifted to the right). 
The results are shown for the central mass and for the standard set of 
$\theta_i$.}
\label{fig:mbare}
\end{figure}

We illustrate the 
cutoff-dependence of $\omega_1=\mhbare$ 
as a function of $L_1/a = L_1\Lambda_\mrm{cutoff}$.
In~\fig{fig:mbare} we show the static bare quark mass and 
its $1/\mbeauty$ contribution in units of $L_1$, using
$L_1/a=L_2/(2a)$ and the data of \tab{table_hqet_param}. 
In addition, smaller lattice spacings 
are covered by including the numbers for $\tilde{\omega}_1$ 
from \eq{e:match}. The two sets of bare parameters $\tilde{\omega}_i$ and ${\omega}_i$
differ by cutoff effects: the former are determined directly for $L=L_1$ and 
the latter after a step scaling to $L=L_2$. 
Indeed these cutoff effects
are visible but not dominating. The largest part
of the variation with $a$ is due to the divergences.
In the static case, the divergence is known
to one-loop order from \cite{stat:actpaper}
(see table~1 of that paper). 
It is plotted for HYP1 and HYP2 using both 
standard (dotted lines) and boosted (dashed lines) 
perturbation theory, fixing the constant piece at the smallest
lattice spacing.
The non-perturbative results agree 
{\it qualitatively} with the perturbative approximations.

\begin{figure}[!htb]
\begin{center}
\begin{tabular}{cc}
\hspace{-0.5cm}
\includegraphics[width=7cm]{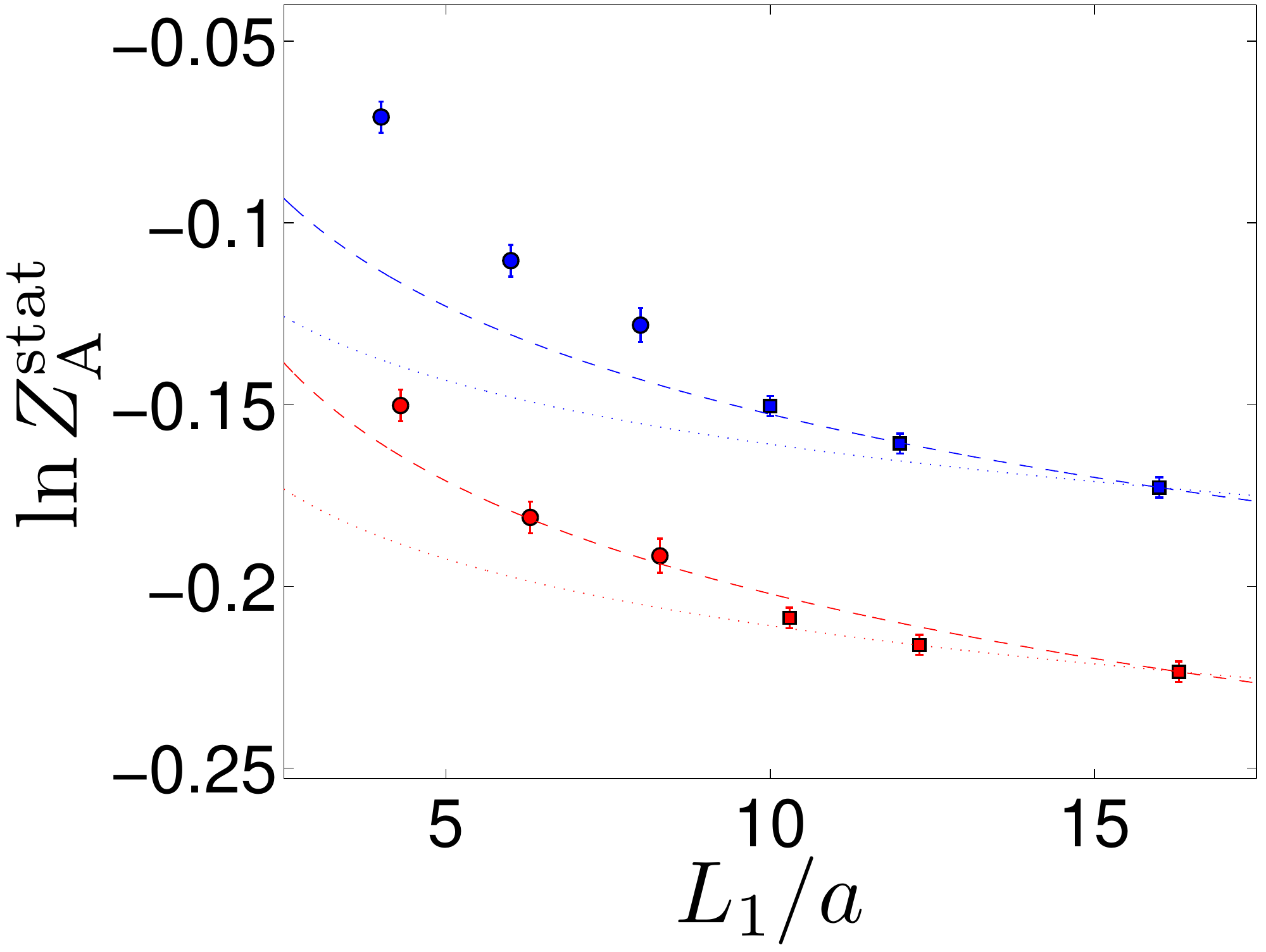}&
\includegraphics[width=7cm]{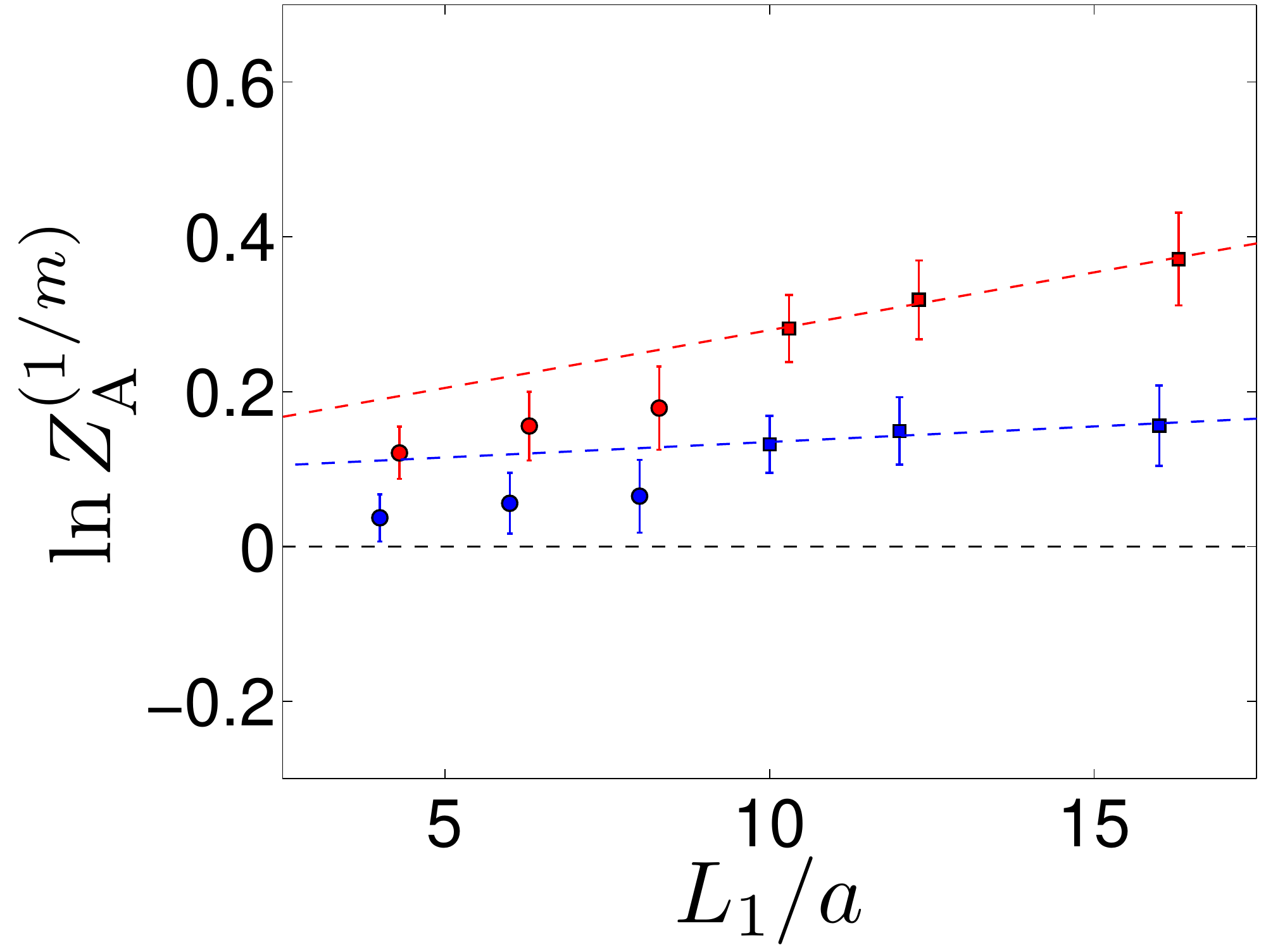}
\end{tabular}
\end{center}
\vspace{-0.7cm}
\caption[]{\footnotesize 
Same as figure \fig{fig:mbare} for $\ln(Z_A)$.}
\label{fig:lnZ}
\end{figure}

We show the equivalent plots for $\lnzastat$ and $\lnzanlo$ 
in~\fig{fig:lnZ}. For $\lnzanlo$ the theoretically expected 
$1/a$ divergence is not clearly visible.  It either
has a rather small coefficient or it is masked by terms
with positive powers of the lattice spacing. For illustration 
we nevertheless fit a linear behaviour to $\lnzanlo$ at 
the smallest three lattice
spacings and extend this curve to larger ones. For $\lnzastat$
we show the leading logarithmic divergence, 
$\lnzastat \sim \ln(a/L_1)\,g_0^2/(4\pi^2)+\rm constant$, in the graph.
Again the constant is adjusted to the data point at smallest $a$
and replacing $g_0^2$ with a boosted coupling defines the 
boosted perturbation theory expression. 

Before closing this section we would like to add a remark concerning 
different matching conditions. 
By construction, the observables $\Phi$ depend on 
the values of the angles $\theta_i$. 
Nevertheless, since the HQET parameters $\omega_i$ are bare parameters 
of the HQET Lagrangian and fields, they are 
$\theta$ independent {\em up to truncation corrections of order
$(1/\mbeauty)^n$}.
As one can see from~\eq{e:phiexp}, this implies that the 
$\theta$-dependence of $\Phi$ is absorbed by the ones of $\eta$ and $\phimat$.
In practice this means that a parameter $\omega$ computed in the static 
approximation, such as $\mhbare^{\stat}$ or $\zastat$ exhibits a small 
$\theta$-dependence, but once the $1/\mbeauty$ corrections are added, 
this dependence
has to be further suppressed to order $(1/\mbeauty)^2$. 
We checked that we obtain this behaviour in our numerical simulations.
It is illustrated for the case of $\omega_4$
in \tab{t:omk}. 
\input{tables/omk_theta.tex}

%% file: tables/simul.tex
\begin{table}[!htb] 
  \hspace{-1.cm} 
  \begin{center} 
    \begin{tabular}{cccc}
      \hline
      Simulation & $L$ & Theory & $L/a$ \\
      \hline
      $S_1$ & $L_1$ & QCD  & $40, 32, 24, 20$     \\
      $S_2$ & $L_1$ & HQET & $16, 12, 10, 8, 6$   \\
      $S_3$ & $L_2$ & HQET & $32, 24, 20, 16, 12$ \\
      $S_4$ & $L_2$ & HQET & $16, 12, 8$          \\
      \hline
    \end{tabular}
  \end{center}
\caption[ ]{  \label{t:simul}
\footnotesize Summary of the simulations used in this work. Note that 
$L_1/a=16,\;L_2/a=32$ are in addition to those of \protect\cite{hqet:pap4}.}
\end{table}

%% file: tables/phiL2stat.tex
 \begin{table}[!htb] 
 \begin{center} 
 \begin{tabular}{ccc}
 \hline
 \hline
 & \multicolumn{2}{c}{$\Phi_i^\stat(L_2,M,0)$}  \\
 \hline
 $z$ & $\Phi_1^\stat$ & $\Phi_2^\stat$ \\
 \hline
 \hline
 $10.4$ & $15.73(20)$  &  $0.538(4)$ \\
 $12.1$ & $17.82(22)$  &  $0.552(4)$ \\
 $13.3$ & $19.29(24)$  &  $0.560(4)$ \\
 \hline \hline
 \end{tabular} 
 \end{center} 
 \caption[ ]{\footnotesize Observables $\Phi_i^\stat(L_2,M,0)$ 
for the different values of the quark mass, 
and for the standard choice of $\theta$'s.}

 \label{t:phiL2stat} 
 \end{table}

%% file: tables/phiL2onem.tex
 \begin{table}[!htb] 
 \begin{center} 
 \begin{tabular}{cccccc}
 \hline
 \hline
 & \multicolumn{5}{c}{$\Phi_i^\first(L_2,M,0)$}  \\
 \hline
 $z$ & $\Phi_1^\first$ & $\Phi_2^\first$ & $\Phi_3^\first$ & $\Phi_4^\first$ & $\Phi_5^\first$\\
 \hline
 \hline
 $10.4$ & $0.14(7)$  &  $0.038(14)$  & $0.0491(23)$  &  $0.0333(4)$  &  $0.00234(16)$ \\
 $12.1$ & $0.12(6)$  &  $0.034(13)$  & $0.0430(22)$  &  $0.0297(4)$  &  $0.00208(14)$ \\
 $13.3$ & $0.11(6)$  &  $0.031(12)$  & $0.0396(21)$  &  $0.0276(4)$  &  $0.00193(13)$ \\
 \hline \hline
 \end{tabular} 
 \end{center} 
 \caption[ ]{\footnotesize Observables $\Phi_i^\first(L_2,M,0)$ 
for the different values of the quark mass, 
and for the standard choice of $\theta$'s.}
 \label{t:phiL2onem} 
 \end{table}

%% file: tables/hqet_param_interp.tex
 \begin{table}[!htb] 
 \hspace{-1.cm} 
 \begin{center} 
 \begin{tabular}{ccccccc}
   \hline
   \hline
   &\multicolumn{3}{c}{HYP1} &\multicolumn{3}{c}{HYP2} \\
 \hline
 $\beta          $  &  6.4956 &  6.2885 &  6.0219  
                    &  6.4956 &  6.2885 &  6.0219 \\ 
 \hline 
 \hline
 $a\mhbare^\stat $  &  $ 0.964(12)$ &  $ 1.324(17)$ &  $ 2.054(25)$   
                    &  $ 0.990(12)$ &  $ 1.352(17)$ &  $ 2.083(25)$ \\ [+0.7ex]
 $\lnzastat      $  &  $-0.182(5)$  &  $-0.171(5)$  &  $-0.141(4)$
                    &  $-0.118(5)$  &  $-0.101(5)$  &  $-0.061(4)$ \\ [+0.7ex]
 \hline

 $a\mhbare^\first$  &  $-0.315(6)$  &  $-0.264(6)$  &  $-0.214(8)$
                    &  $-0.328(6)$  &  $-0.273(6)$  &  $-0.215(8)$ \\ [+0.7ex]
 $\lnzanlo       $  &  $ 0.180(28)$ &  $ 0.156(24)$ &  $ 0.121(20)$   
                    &  $ 0.068(27)$ &  $ 0.058(24)$ &  $ 0.039(20)$ \\ [+0.7ex]
 $\cahqet/a      $  &  $-0.17(7)$   &  $-0.06(5)$   &  $ 0.03(4)$
                    &  $-0.61(7)$   &  $-0.46(5)$   &  $-0.28(4)$ \\ [+0.7ex]
 $\omegakin/a    $  &  $ 0.550(9)$  &  $ 0.437(7)$  &  $ 0.328(5)$   
                    &  $ 0.553(9)$  &  $ 0.439(7)$  &  $ 0.330(5)$ \\ [+0.7ex]
 $\omegaspin/a   $  &  $ 0.76(5)$   &  $ 0.59(4)$   &  $ 0.43(3)$   
                    &  $ 0.87(6)$   &  $ 0.71(5)$   &  $ 0.55(4)$ \\ [+0.7ex]
 \hline \hline
 \end{tabular} 
 \end{center} 
 \caption[ ]{\footnotesize HQET parameters as a function of the bare coupling
 for the actions HYP1 and HYP2, for our choice of $\theta_i$ and for 
 $z=\Mbeauty L=12.30$ (static) or $z=\Mbeauty L=12.48$.
 The three values of $\beta$ correspond to $L_2/a=16,12,8$ respectively.}
 \label{table_hqet_param} 
 \end{table}

%% file: tables/omk_theta.tex
\begin{table}[!htb] 
 \hspace{-1.cm} 
 \begin{center} 
 \begin{tabular}{ccccc}
 \hline
 \hline
& $L_2/a$  & $(\theta_1, \theta_2) = (0,0.5)$ & $(\theta_1, \theta_2) = (0.5,1)$ & $(\theta_1, \theta_2) = (0,1)$ \\
 \hline
 \hline
$\Phi_4^{\first}$ &  & $0.0098(3)$ &  $0.0297(4)$  &  $0.0395(7)$ \\
\hline
$\omegakin/a$& $16$ &$0.566(18)$  &  $0.565(7)$  &  $0.566(10)$ \\
$\omegakin/a$& $12$ &$0.450(15)$  &  $0.449(6)$  &  $0.450(8)$  \\
$\omegakin/a$& $8$  &$0.339(11)$  &  $0.338(4)$  &  $0.338(6)$  \\
 \hline 
 \hline
 \end{tabular} 
 \end{center} 
 \caption[ ]{\footnotesize $\theta$-- dependence of $\Phi_4^{\first}(L_2,M,0)$ 
and  the corresponding HQET parameter 
$\omega_4(M,a)=\omegakin(M,a)$, for $z=12.1$ and the action HYP2.}
 \label{t:omk} 
 \end{table}

%% file: s5.tex
\section{Outlook}

Our non-perturbatively computed HQET parameters show a qualitative
agreement with perturbation theory as far as the leading divergence 
in each of the parameters is concerned (see figures~\ref{fig:mbare},\ref{fig:lnZ}),
but this does not extend to the quantitative
level needed for precision flavour physics.

Fortunately, in the quenched approximation,
we now have the full set of parameters for HQET spectrum calculations
as well as for (zero space-momentum) matrix element of $A_0$ including 
the terms of order $\minv$ -- all of them are known non-perturbatively.
A detailed test of the $\minv$ expansion is thus possible. 
In companion papers we are carrying this out for the examples of
the $\rm B_s$ decay constant and for some mass splittings in the 
$\rm B_s$ system; some preliminary results are described
in \cite{lat09:tereza}. 
The parameters do of course depend
on our choice of discretization, but starting from 
Tables~\ref{t:phiL2stat} and \ref{t:phiL2onem}, the effort to 
compute them for a different regularization is very modest.
Note that,  as long as one remains  
in the quenched approximation,
the parameters for the HQET action do not need to be 
recomputed if one uses just a 
light quark action differing from ours.

The small volume simulations for a determination of the parameters
for $N_f=2$ are already far advanced, see \cite{lat08:patrick}
for a recent account. Therefore, the situation will soon be similar
with 2 flavours of dynamical fermions.

%% file: a1.tex
\section{Choice of observables
\label{a:choice}}

Our observables are built from \SF 
\cite{SF:LNWW,SF:stefan1} correlation functions
defined exactly as in \cite{hqet:pap4}. We use
the renormalized boundary-to-bulk correlation function $\fa(x_0,\theta)$
in the pseudoscalar channel as well as the boundary-to-boundary 
correlators $\fone,\,\kone$, in 
pseudoscalar and vector channel respectively.
For their interpretation, we note in particular the r\^ole 
of $\theta$. Quark fields $\psi$ are periodic in space
up to a phase $\rme^{i\theta}$, while $\psibar$ are periodic 
up to $\rme^{-i\theta}$. This shifts the lowest momentum of a free
quark from $\vecp=0$ to $\vecp=(1,1,1)\theta/L$.
Intermediate states to the above correlation functions
have vanishing momentum, due to a projection $a^3\sum_\vecx$
and the fact that fields $\psibar_1 \Gamma \psi_2$ are
periodic (with a vanishing phase). However, in the free
theory, which is a relevant starting point for the interpretation
in small volume, the quark carries momentum 
$\vecp=(1,1,1)\theta/L$, compensated by the opposite momentum of
the anti-quark.

We form an effective energy 
$\meffp = -\left.\tilde\partial_0(\ln(-\fa(x_0,\theta_0))\right|_{(x_0=T/2,T=L)}$ 
which approaches the heavy quark mass as it becomes very large.
The logarithm of the ratio 
$\ra=\left.\ln(\fa(T/2,\theta_1)/\fa(T/2,\theta_2)\right|_{T=L}$\,
is easily seen to have a sensitivity to the coefficient $\omega_3$ 
which is approximately proportional to $\theta_2-\theta_1$: one just has
to note that 
the covariant derivatives in \eq{e:dahqet} 
acting on the quark fields are proportional to their 
momentum. This free theory argument is valid qualitatively 
in small volume. In the same way one sees that the combination
$\rone=\frac14\left(
\ln(\fone(\theta_1)\kone(\theta_1)^3)-\ln(\fone(\theta_2)\kone(\theta_2)^3)
\right)$
has a sensitivity proportional to $\theta_2^2-\theta_1^2$ to $\omega_4$,
while $\omega_5$ does not contribute due to spin symmetry.

These qualitative considerations combined with some
numerical experiments lead us to introduce
\bes
\Phi =     \big(L\meffp 
      \,,\; \ln\left({-\fa\over\sqrt{\fone}}\right)
      \,,\; \ra
      \,,\; \rone
      \,,\; {3\over4}\ln\left({\fone\over\kone}\right)
\big)^\mrm{t}\,.
\ees

%% file: a2.tex
 \section{Explicit form of step scaling functions
\label{a:ssf}}
From our definitions we find immediately
\bes
  &&\Phi(L_2,M,0) =
  \\ && \nonumber
  \lim_{a/L_1\to0}\left\{\phistat(L_2,a) +
  \phimat(L_2,a)\phimat^{-1}(L_1,a)[\Phi(L_1,M,0)-\phistat(L_1,a)]
  \right\}\,,
\ees
where 
\bes
  \phimat(L_2,a)\phimat^{-1}(L_1,a) &=&
  \pmat{D & [B(L_2,a)-D B(L_1,a)] \,A^{-1}(L_1,a) \\ 0
  & A(L_2,a)A^{-1}(L_1,a)}\,, \;\\ D&=&\diag(L_2/L_1,1)
\ees
in terms of the matrices introduced in \eq{e:AB}. The form
\bes
  \Phi_i(L_2,M,0) &=& D_i\, \Phi_i(L_1,M,0) +
                      \lim_{a/L_1\to0}\,\widehat\Sigma_{i}(L_1,a)
                   \\ && \nonumber +
                      \lim_{a/L_1\to0} \sum_{j=3}^5\, \Sigma_{ij}(L_1,a)
                      [\Phi_j(L_1,M,0)-\phistat_j(L_1,0)] \,,\;i=1,2\,,
  \\
  \Phi_i(L_2,M,0) &=&\phistat_i(L_2,0) +
                      \lim_{a/L_1\to0} \sum_{j=3}^5\Sigma_{ij}(L_1,M,a)
                      [\Phi_j(L_1,M,0)-\phistat_j(L_1,0)] \,,
  \nonumber \\[-1ex] &&\qquad\qquad\qquad\qquad\qquad\qquad\;i=3\ldots5\,.
\ees
then introduces step scaling functions
\bes
  \label{e:siga}
  \widehat\Sigma_{i}(L_1,a)
  &=& \phistat_i(L_2,a) - D_i\,\phistat_i(L_1,a)\,, \;i=1,2\,,
  \\
  \Sigma(L_1,a) &=& \pmat{D &
                     [B(L_2,a)-D\,B(L_1,a)]A^{-1}(L_1,a)
                         \\ 0 &  A(L_2,a)\,A^{-1}(L_1,a)} \,.
\ees
By choosing only $L_1$ and $a$ as arguments we have assumed
that $L_2/L_1=s$ is fixed (typically to $s=2$) which means
$D=\diag(s,1)$.

The continuum limit of each element of the step scaling functions
exists and the above split into two
pieces is suggested by the fact that the limit
$a\to0$ of $\phistat_i(L,M,a)$ may be performed  for $i\geq3$,
while for $i\leq2$ there is an additive renormalization
which only cancels in \eq{e:siga}. Splitting $\phistat$ accordingly,
$\phistat=\phistat^a + \phistat^b$ with $\phistat^a_{i\geq3}=0\,,\;
\phistat^b_{i\leq2}=0$, we can also unify the step from
$L_1$ to $L_2$ to (note $\widehat\Sigma(L,a)_{i\geq3}=0$)
\bes
   \label{e:evol}
   \Phi(L_2,M,0) = \phistat^b(L_2,0) + \lim_{a/L_1\to0} \, \Sigma(L_1,a)
                      [\Phi(L_1,M,0)-\phistat^b(L_1,0)] + \widehat\Sigma(L_1,a)\,.
\ees

Let us now list explicit expressions for the various matrices.
The expansion of the observables in HQET follows directly from
the expansions detailed in section 2.3 of \cite{hqet:pap4}.
The inhomogeneous part is given by the static quantities
\bes
   \phistat= \big(L\meffstat
      \,,\; \zetaa
      \,,\; \rastat
      \,,\; \ronestat
      \,,\; 0 \big)^\mrm{t}\,,
\ees
where 
\bea
\meffstat(L) &=& \left. - {{\partial_0 + \partial_0^* }\over {2}} 
(\ln (-\fastat(x_0,\theta_0)))\right|_{(x_0=T/2,T=L)} 
\,,\\
\zetaa(L) &=& \left.\ln\left(
{- \fastat (x_0,\theta_0)\over {\sqrt{\fonestat(\theta_0)}}}
\right)\right|_{(x_0=T/2,T=L)} 
\,,\\
\rastat(L) &=& \left.\ln\left(
{\fastat(x_0,\theta_1) \over \fastat(x_0,\theta_2)}
\right)\right|_{(x_0=T/2,T=L)} 
\,,\\
\ronestat(L) &=& \left.\ln\left(
{\fonestat(\theta_1) \over \fonestat(\theta_2)}
\right)\right|_{T=L/2} 
\,.
\eea 
Note that $\fastat$ does not contain the improvement term proportional
to $\castat$. The improvement term is included as explained in
section~\ref{s:param}. 
From the definition eq.~(\ref{e:siga}), one has 
\bes 
\label{e:Sig1}
\widehat\Sigma_1 &=& L_2(\meffstat(L_2)-\meffstat(L_1)) \,,\\
\label{e:Sig2}
\widehat\Sigma_2 &=& \zetaa(L_2)-\zetaa(L_1))\,.
\ees
The matrices $A,B$ which make up $\phimat$ as in
\eq{e:AB} are
\bes
  A=\pmat{ \rda & \rakin   & \raspin \\
          0 &\ronekin & 0 \\
          0 & 0 &  \rhoonespin} \quad
  B=\pmat{ L\meffdastat & L\meffkin & L\meffspin \\
           \rhod &  \psikin & \psispin}\,,
\ees
where
\bea
\meffy(L) &=&  - \left.{{\partial_0 + \partial_0^* }\over {2}} 
(\fay(x_0,\theta_0)/\fastat(x_0,\theta_0))\right|_{(x_0=T/2,T=L)}
\,,\\
\meffdastat(L) &=&  - \left.{{\partial_0 + \partial_0^* }\over {2}} 
(\fdeltaastat(x_0,\theta_0)/\fastat(x_0,\theta_0))\right|_{(x_0=T/2,T=L)}
\,,\\
\psiy(L)  &=& 
\left.
\left( {\fay(x_0,\theta_0) \over \fastat(x_0,\theta_0)}
- {1\over2} {\foney(\theta_0) \over \fonestat(\theta_0)} \right)
\right|_{(x_0=T/2,T=L)}
\,,\\
\rho_1^{\rm y}(L) &=& \left.{
\foney(\theta_0) \over \fonestat(\theta_0)}\right|_{T=L/2} 
\,,\\
\rhod(L) &=& \left.{
\fdeltaastat(x_0,\theta_0) \over \fastat(x_0,\theta_0)}\right|_{(x_0=T/2,T=L)}
\,,\\
\ray(L) &=& \left.\left(
{\fay(x_0,\theta_1) \over \fastat(x_0,\theta_1)} - {\fay(x_0,\theta_2) \over \fastat(x_0,\theta_2)}
\right)\right|_{(x_0=T/2,T=L)} 
\,,\\
\rda(L) &=& \left.\left(
{\fdeltaastat(x_0,\theta_1) \over \fastat(x_0,\theta_1)} - {\fdeltaastat(x_0,\theta_2) \over \fastat(x_0,\theta_2)}
\right)\right|_{(x_0=T/2,T=L)} 
\,,\\
\roney(L) &=& \left.\left(
{\foney(\theta_1) \over \fonestat(\theta_1)} - {\foney(\theta_2) \over \fonestat(\theta_2)}
\right)\right|_{T=L/2} \,.
\eea
In order to avoid unnecessary repetitions,  we have used 
the shortcut
\bes
\mrm{y} \in \{\mrm{kin}, \mrm{spin} \} \,.
\ees
in the previous formulae. From our definitions we obtain
\bes
  A^{-1}=\pmat{ 1/\rda & K_{12}   & K_{13} \\
          0 & 1/\ronekin & 0 \\
          0 & 0  & 1/\rhoonespin}\,,\quad \\
          K_{12} = {-\rakin \over \ronekin \rda}\,,\;
          K_{13} = {-\raspin \over \rhoonespin \rda }\,,
\ees
and
\bes
 [B(L_2,a)-D\,B(L_1,a)]A^{-1}(L_1,a)=
         \pmat{\Sigma_{13} & \Sigma_{14} & \Sigma_{15} \\
                \Sigma_{23}& \Sigma_{24} & \Sigma_{25} \\
               }\,,\\
 A(L_2,a)\,A^{-1}(L_1,a) = \pmat{
               \Sigma_{33} & \Sigma_{34} & \Sigma_{35} \\
                0 & \Sigma_{44} & 0 \\
                0 & 0           & \Sigma_{55} 
               }\,,
\ees
with
\bes
   \nonumber \Sigma_{13} &=& L_2(\meffdastat(L_2)-\meffdastat(L_1))/\rda(L_1) \,,\\
   \nonumber \Sigma_{14} &=& L_2(\meffdastat(L_2)-\meffdastat(L_1))\,K_{12}(L_1)
                  +L_2(\meffkin(L_2)-\meffkin(L_1))/\ronekin(L_1)\,,\\
   \nonumber \Sigma_{15} &=& L_2(\meffdastat(L_2)-\meffdastat(L_1))\,K_{13}(L_1) 
                  +L_2(\meffspin(L_2)-\meffspin(L_1))/\rhoonespin(L_1)\,,\\
   \nonumber \Sigma_{23} &=& (\rhod(L_2)-\rhod(L_1))/\rda(L_1)\,, \\
   \nonumber \Sigma_{24} &=& (\rhod(L_2)-\rhod(L_1))\,K_{12}(L_1) 
                  +(\psikin(L_2)-\psikin(L_1))/\ronekin(L_1)\,,\\
   \nonumber \Sigma_{25} &=& (\rhod(L_2)-\rhod(L_1))\,K_{13}(L_1) 
                  +(\psispin(L_2)-\psispin(L_1))/\rhoonespin(L_1)\,,\\
 \ees
and
\bes
   \nonumber \Sigma_{33} &=& \rda(L_2)/\rda(L_1) \,,\\
   \nonumber \Sigma_{34} &=&
                   \rda(L_2)K_{12}(L_1)+\rakin(L_2)/\ronekin(L_1) \,,\\
   \nonumber \Sigma_{35} &=&
                   \rda(L_2)K_{13}(L_1)+\raspin(L_2)/\rhoonespin(L_1) \,,\\
   \nonumber \Sigma_{44} &=& \ronekin(L_2)/\ronekin(L_1) \,,\\
   \nonumber \Sigma_{55} &=& \rhoonespin(L_2)/\ronespin(L_1) \,.\\
\ees
In static approximation the only non-vanishing elements
are
\bes
   \nonumber \Sigma_{13}^\stat = \Sigma_{13}\,,\quad 
   \nonumber \Sigma_{23}^\stat = \Sigma_{23}\,.
\ees

%% file: a3.tex
\section{Covariance matrices
\label{a:cov}}
When our parameters are used in a subsequent computation,
one has to note that their errors are correlated. We therefore
supply covariance sub-matrices as far as they appear necessary in practice.
They are still too large to be printed (and typed in). We therefore
supply them for download together with the paper under 
{\tt http://arxiv.org/}
and at
{\tt http://www-zeuthen.desy.de/alpha/public\_tables/}. All parameters are given in lattice units $a=1$, and
for our standard choice of $\theta$ angles. 
As explained in the text, we give the HQET parameters and their covariance 
matrices for $\castat$ fixed to its one-loop value~\cite{castat:filippo}
and fixed to 0.

We normalize the covariance matrix between two quantities $O_\alpha$ and
$O_\beta$ in the following way
\bes\nonumber
C_{\alpha \beta} &=& {{\sigma_{\alpha \beta}} \over {\sigma_\alpha \sigma_\beta}}\,, \\
\sigma_{\alpha \beta} &=& (N_{\rm jack}-1) 
( \langle O_\alpha O_\beta \rangle - \langle O_\alpha\rangle \langle O_\beta \rangle )
\,,
\ees
where $\langle \,\rangle$ represents the average over $N_{\rm jack}=100$ jackknife samples.
The covariance matrix is such that $C_{\beta \alpha} = C_{\alpha \beta}$ 
and $C_{\alpha \alpha}=1$ (no summation).\\
A first subset of parameters is 
$ (\omega_1^\stat,\,  \omega_1^\first,\, \omega_4)$, 
necessary for the computation of the quark mass.
In order to take into account the different kinds of correlations 
(correlation between different values of the lattice spacing, 
or between different values of the heavy quark mass),
the indices $\alpha$ and $\beta$ take the values 
\bes
\alpha,\beta = l + 3(i-1) +9(j-1) +27(k-1) \,.
\ees
Here $l=1,2,3$ represents the parameters $\omega_1^\stat,\omega_1^\first$ 
and $\omega_4=\omega_4^\first$ respectively, $i=1,2,3$ labels the three $g_0$ values, 
$j=1,2,3$ numbers the three values of the quark mass and $k=1(2)$ 
for the HYP1(2) static actions. 
This leads to a $54\times54$ matrix $C_{\alpha\beta}^{a}$, but for symmetry 
reasons, we give only the $1431$ components with $\alpha < \beta$.

The second subset is the one relevant for a computation 
of the pseudoscalar decay constant.
There the indices $\alpha$ and $\beta$ take the value
\bes
\alpha,\beta = l+5(i-1) +15(k-1) \,.
\ees
Here 
$l = 1$ represents $\omega_{2}^\stat$ and
$l=2\ldots 5$ represent $\omega_{2}^\first, \ldots,\omega_{5}^\first$. 
while $i,k$ have the same meaning as before.
Thus we obtain a $30\times 30$ matrix $C_{\alpha\beta}^{b}$, and give 
its $435$ relevant components.
The heavy quark mass is fixed  by matching to the spin averaged 
$\rm B_s$-meson mass. This matching is done at the static order for 
$\omega^\stat$ and including the $\minv$ terms for $\omega^\first$. 

%% file: a4.tex
\section{Tree level improvement}
\label{a:tli}

Perturbative improvement of the observables \cite{alpha:SU2impr}
has been proven to be effective in 
reducing cutoff effects for example in the case of the step scaling function 
for the coupling~\cite{pert:2loop_fin,alpha:nf2}, 
where it has been pushed to the 
two-loop order. The idea is to remove all the 
O(($\frac{a}{L})^n\ln(\frac{a}{L})^m)$ at a given order in 
the $\bar{g}^2$  expansion of the renormalized lattice observable. 
At tree level, as done here, this produces
{\it classically perfect} observables in the sense of~\cite{Hasenfratz:1998bb},
by removing all the O($(\frac{a}{L})^n$) effects.

The tree-level improved observable is either defined as 
\be 
  \label{e:multimpr}
  O_\mrm{impr}\left({a/ L}\right) = {O\left({a/ L}\right) \over 1 + \delta\left({a/ L}\right) }\,,
  \quad
\delta\left({a/ L}\right)=
{{O^{\rm tree}\left({a/ L}\right)-O^{\rm tree}(0)}\over{O^{\rm tree}(0)}}\;,
\ee 
or as 
\be
  \label{e:addimpr}
  O_\mrm{impr}\left({a/ L}\right) = O\left({a/ L}\right) - \delta\left({a/ L}\right)\,,
  \quad
\delta\left({a/ L}\right)=O^{\rm tree}({a/ L})-O^{\rm tree}(0)\;.  
\ee
Here $O$ is the non-perturbative observable evaluated 
by Monte Carlo, $O_\mrm{impr}$ is the tree-level improved one
and $O^{\rm tree}$ is the same observable evaluated for $g_0=0$.
Generically we use \eq{e:multimpr}, but when $O^{\rm tree}(0)$ vanishes
\eq{e:addimpr} is appropriate. We also choose the latter for
the step scaling functions $\widehat{\Sigma}$ since it is
more natural given their form \eq{e:Sig1}, \eq{e:Sig2}.
We apply this improvement to all QCD and HQET observables used for
the matching in small volume and to all HQET step scaling
functions.

At tree level there are some useful relations among the Schr\"odinger
Functional  correlation 
functions, which can be used in the perturbative computation~\cite{impr:pap5}:
\bea
k_{\rm V}(x_0)&=&{2\over3}f_{\rm P}(x_0)-{1\over3}f_{\rm A}(x_0) \;, \\
f_1&=& k_1\;=\;\left\{ {1\over2}f_{\rm P}(x_0)-{1\over2}f_{\rm A}(x_0) 
\right\}_{x_0=T-a} \;.
\eea
Since in our QCD simulations
$L$ is fixed through $\gbar^2(L)$, each 
value of $z$ determines a value of $M$. In the corresponding {\em tree level} computation
the improved subtracted quark mass~\footnote{Here 
$\bm$ is an improvement coefficient \cite{impr:pap1}.} 
$\mqtilde=\mq(1+\bm a\mq)$ enters. 
At the considered order we could set
$\mqtilde$   to $\mbar(\mu)$ at an arbitrary scale 
$\mu$ in an arbitrary scheme. 
We decided to choose $\mqtilde=m_\star$, where $m_\star \equiv \mbar_\mrm{univ}(m_\star)$
and $\mbar_\mrm{univ}(\mu)$ is the running mass computed from
the RGI mass (but) using only 
the universal part of the $\beta$ (i.e. the coefficients $b_0$ and
$b_1$) and $\tau$ (i.e. the coefficient $d_0$) functions. This choice is 
scheme independent and it is based on the expectation that cutoff effects are
dominated by scales around $\mu=m_\star$. The explicit
relation between $z$ and $m_\star$, 
\be
{\Lambda_{\rm SF}L\over{z}}=2^{b_1/(2b_0^2)}\; x^{1-b_1/(b_0d_0)}\;
e^{-x^{-2b_0/d_0}} \quad {\rm with} \quad
x={m_\star\over{M}}\,.
\ee
is then solved for $m_\star$. 
In our non-perturbative computation,
the $\Phiqcd_i$ 
have been computed for $z=ML= 10.4,\; 12.1\; 
{\rm and} \;13.3$ 
with $\gbar^2(L)=\tilde u_1$ 
(see the beginning  of \sect{s:num}) 
and correspondingly $\Lambda_{\rm SF}L=0.195(16)$. Our tables,
provided under  \texttt{http://www-zeuthen.desy.de/alpha/public\_tables/},
are for these specific values. 

\begin{figure}[!htb]
\begin{center}
\begin{tabular}{cc}
\hspace{-0.6cm}
\includegraphics[width=7cm]{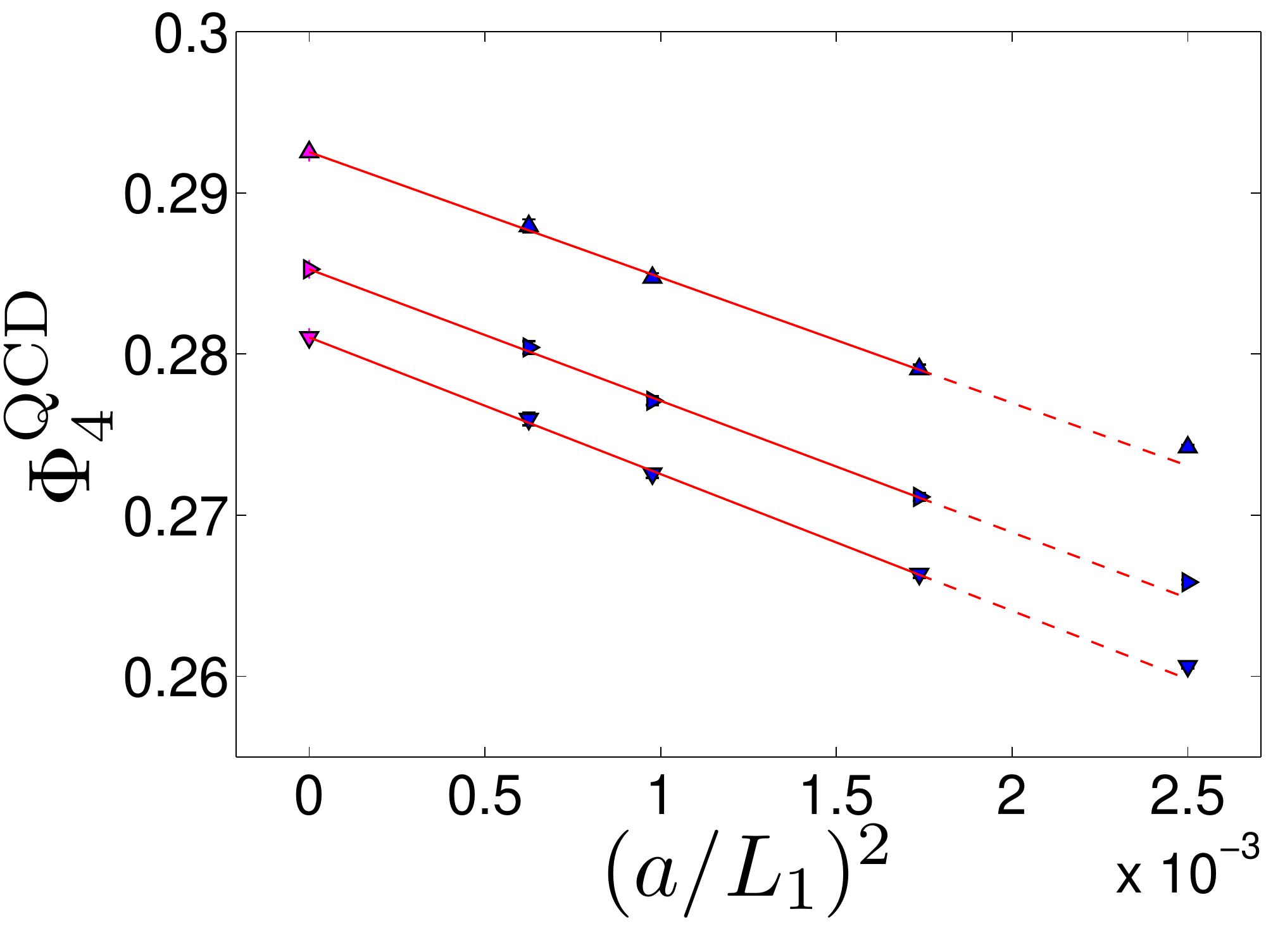}&
\includegraphics[width=7cm]{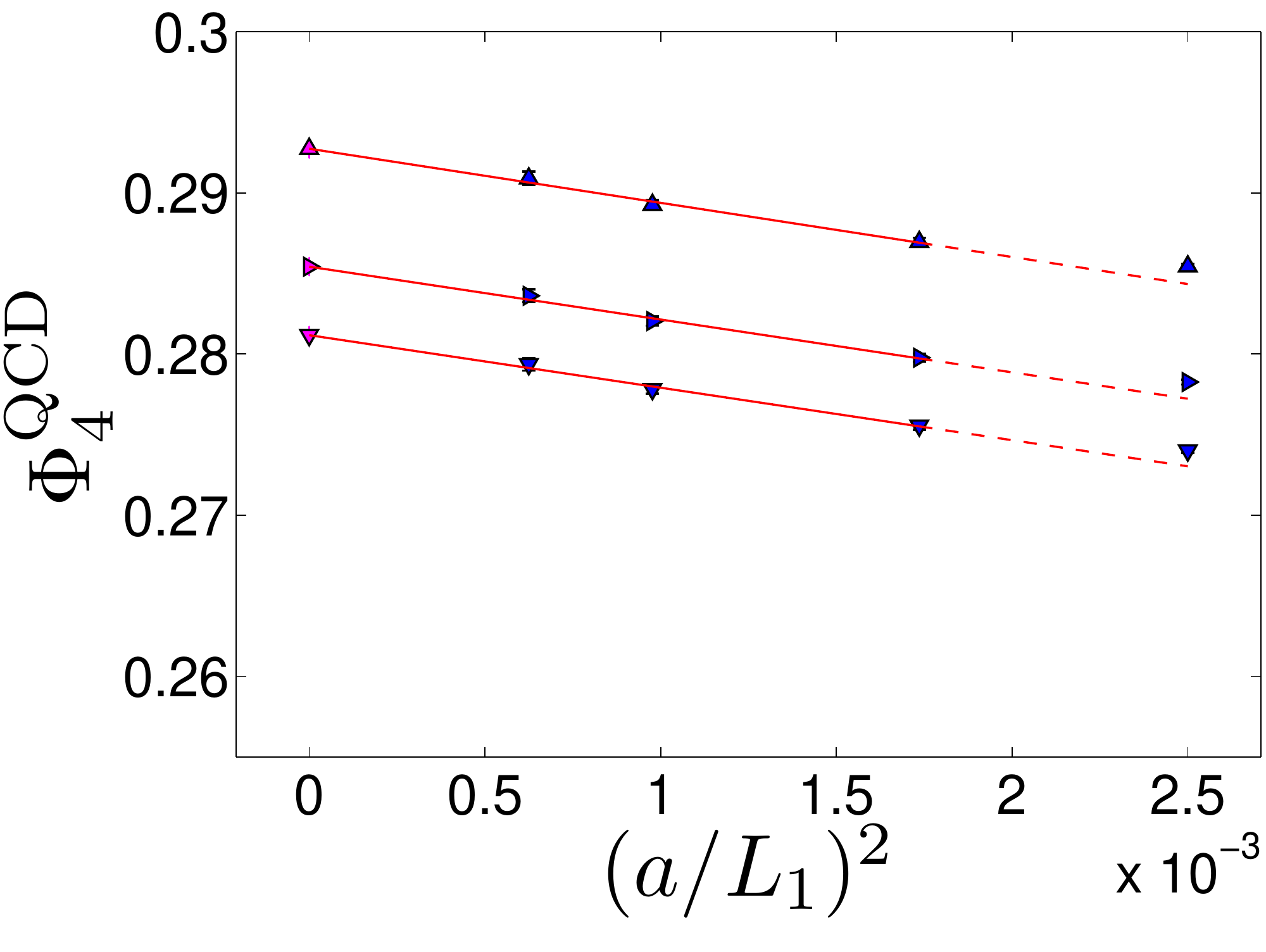}\\
\end{tabular}
\end{center}
\vspace{-0.7cm}
\caption[]{\footnotesize 
Effect of tree level improvement on $\Phi^\mrm{QCD}_4$.
The three masses are shown,
with the same conventions as in fig~\ref{fig:Phi1}.}
\label{fig:1appD}
\end{figure}

As an example, 
the non-perturbative $\Phiqcd_4$ with (right) and without
 (left) tree level improvement of the observable are compared for 
$(\theta_1,\theta_2)=(0.5,1)$ in \fig{fig:1appD}. 

Turning now to HQET, the correlation 
functions are defined in~\cite{hqet:pap4} and useful relations at tree level are
\bea
f_{\rm A}^{\rm kin}(x_0) &=& 
6 {x_0 \over a^2}\left(\cos\left( {{a\theta}\over{L}}\right)-1\right) 
f_{\rm A}^{\rm stat}(x_0) \;, \\
f_1^{\rm kin}&=& k_1^{\rm kin}\;=\;
6{(T-a) \over a^2}\left(\cos\left( {{a\theta}\over{L}}\right)-1\right) 
f_1^{\rm stat} \;,  \\
f_{X}^{\rm spin}&=&0 \;.
\label{e:spintree}
\eea
The last relation holds when the magnetic background field vanishes,
which is the case in our application.

\begin{figure}[!htb]
\begin{center}
\begin{tabular}{cc}
\hspace{-0.6cm}
\includegraphics[width=7cm]{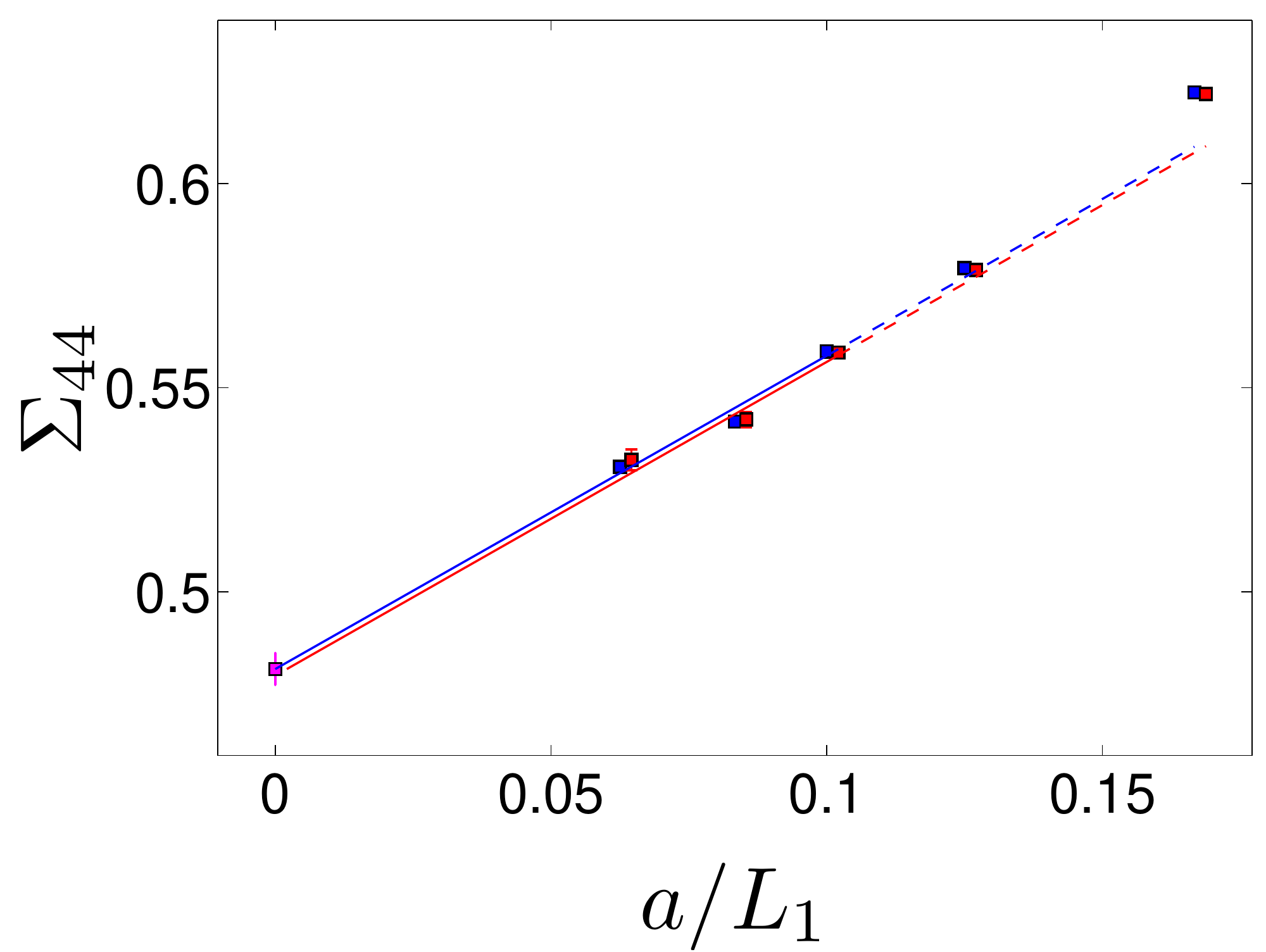}&
\includegraphics[width=7cm]{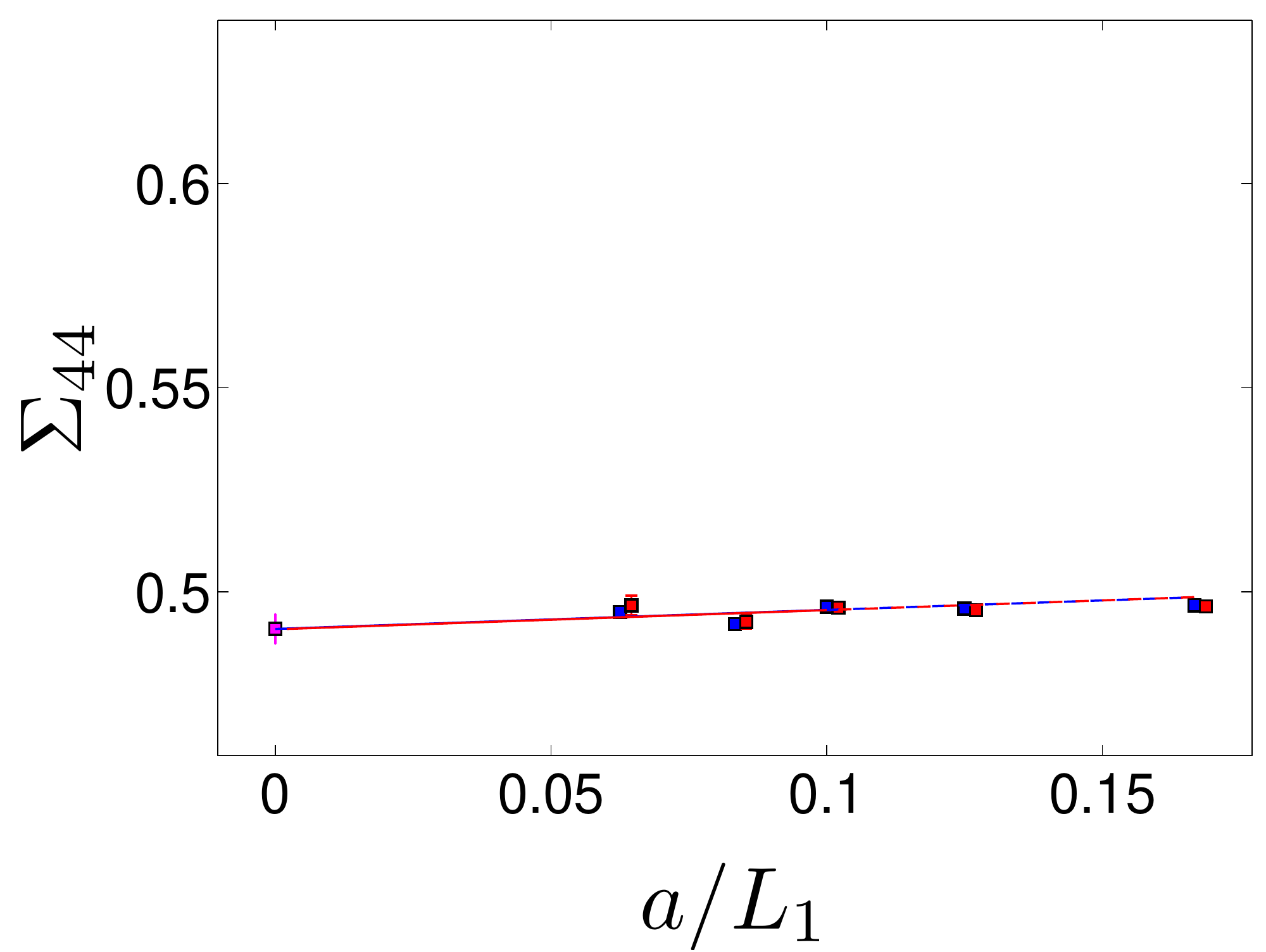}\\
\end{tabular}
\end{center}
\vspace{-0.7cm}
\caption{\footnotesize 
Effect of tree level improvement on $\Sigma_{44}$
for $(\theta_1, \theta_2) = (0.5, 1)$. 
The conventions are the same as in fig~\ref{fig:etab}.}
\label{fig2:appD}
\end{figure}

It is instructive to consider now the step scaling function
$\Sigma_{44}$, which has also been discussed 
in~\cite{hqet:pap4}\footnote{There it is  called $\Sigma_{\rm kin}$.},
but without tree level improvement.
Explicitly
\be
\Sigma_{44} (u,a/L)=[R_1^{\rm kin}(2L)/R_1^{\rm kin}(L)]_{\bar{g}^2(L)=u}\,,
\ee
with
\be
R_1^{\rm kin}(L,\theta_1,\theta_2)=f_1^{\rm kin}(\theta_1)/f_1^{\rm stat}
(\theta_1)- f_1^{\rm kin}(\theta_2)/f_1^{\rm stat}(\theta_2) \;, \quad T=L/2\;.
\ee
Given the tree level relations among correlation functions, 
the expression for the step scaling function reads
\be
\Sigma_{44}^{\rm tree}(a/L)={{{L\over a}-1}\over{{L\over 2a}-1}} \cdot 
{ {\left( \cos\left({{a\theta_1}\over{2L}}\right)-  
\cos\left( {{a\theta_2}\over{2L}}\right) \right)} 
\over{\left(   \cos\left({{a\theta_1}\over{L}}\right)-  
\cos\left({{a\theta_2}\over{L}}\right) \right)}  }\;,
\ee
with $\Sigma_{44}^{\rm tree}(0)=0.5$~. 
The first factor is responsible for cutoff effects linear in $a$,
which appear because we have not taken care to $\rmO(a)$ improve
the theory at order $\minv$. 
The correction $\delta$ is then large and tree-level improvement 
leads to a strong reduction of the cutoff effects in the 
non-perturbative $\Sigma_{44}$, as illustrated
in \fig{fig2:appD}.

For future applications, this discussion suggests to 
eliminate linear (tree level) $a/L$ effects such as
those in \fig{fig2:appD} from the start. This is achieved by
implementing $\rmO(a)$-improvement at tree-level
in the order $\minv$ terms of the action and the effective fields. 
It just means to add terms
\bes
   \delta S &=& - a^3 \,\sum_\vecx 
   [{1\over2} \Okin(0,\vecx) + {1\over2} \Okin(T,\vecx)] \,,\\
   \Ah{3} &=& - {1\over2} \lightb(x)\gamma_0\gamma_5 \vecD^2 \heavy(x)\,,
\ees
to action and currents. The normalization is chosen such that
a coefficient $\omegakin$ will implement tree-level improvement,
\bes
 \Ahqet(x)&=& \zahqet\,[\Astat(x)+ \cah{1}\Ah{1}+ \cah{2}\Ah{2}
              + a\omegakin \Ah{3}]\,,
\ees
and similarly for the action. The effect of these terms
is that the kinetic operator $\vecD^2$ is inserted only with weight 
$1/2$ on the initial and final time-slice of the correlation functions
such as \eq{e:caakinr} or $\fakin,\fonekin,\konekin$. After the Wick-contraction
(in a given gauge background field) this corresponds 
to a standard discretized representation of the integral over the time
position of the insertion of $\vecD^2$ in the static propagator.
The spin operator $\Ospin$ can be treated in complete analogy, but
due to \eq{e:spintree} this has no effect at tree-level.

We provide the tree-level improvement coefficients under \\
\texttt{http://www-zeuthen.desy.de/alpha/public\_tables/}.